\documentclass[twocolumn,showpacs,floats,floatfix,superscriptaddress,aps,pra]{revtex4-1}
\usepackage{amsfonts,amssymb,amsmath}
\usepackage{color,calc}
\usepackage[dvips]{graphicx}
\usepackage{bm}
\usepackage[normalem]{ulem}

\def\be{ \begin{equation} }
\def\ee{ \end{equation} }
\def\bea{ \begin{eqnarray} }
\def\eea{ \end{eqnarray} }
\def\bse{ \begin{subequations} }
\def\ese{ \end{subequations} }

\def\i{\,\text{i}}
\def\e{\,\text{e}}
\def\i{i}
\def\e{e}

\def\to{\rightarrow}

\def\d{\text{d}}

\def\T{\mathcal{T}} % time-ordering operator

\def\black{}

\def\to{\rightarrow}

\def\i{\text{i}}

\begin{document}

\author{Genko T. Genov}
\affiliation{Institut for Quantum Optics, Ulm University, Albert-Einstein-Allee 11, Ulm 89081, Germany}

\author{Nati Aharon}
\affiliation{Racah Institute of Physics, The Hebrew University of Jerusalem, Jerusalem 91904, Givat Ram, Israel}

\author{Fedor Jelezko}
\affiliation{Institut for Quantum Optics, Ulm University, Albert-Einstein-Allee 11, Ulm 89081, Germany}

\author{Alex Retzker}
\affiliation{Racah Institute of Physics, The Hebrew University of Jerusalem, Jerusalem 91904, Givat Ram, Israel}

\title{Mixed Dynamical Decoupling}

\date{\today}

\begin{abstract}
We propose a scheme for mixed dynamical decoupling (MDD), where we combine continuous dynamical decoupling with robust sequences of phased pulses. Specifically, we use two fields for decoupling, where the first continuous driving field creates dressed states that are robust to environmental noise. Then, a second field implements a robust sequence of phased pulses to perform inversions of the dressed qubits, thus achieving robustness to amplitude fluctuations of both fields. We show that MDD outperforms standard concatenated continuous dynamical decoupling in realistic numerical simulations for dynamical decoupling in NV centers in diamond. Finally, we also demonstrate how our technique can be utilized for improved sensing.
\end{abstract}

\maketitle

%%%%%%%%%%%%%%%%%%%%%%%%%%%%%%%%%%%%%%%%%%%%%%%%%%%%%%%%%%%%%%%%%%%%%%%%%%%%%%%%%%%%%%%%%%%%%%%%%%%%%%%%%%%%%%%%%%%%%%%%%%%%%%%
%%%%%%%%%%%%%%%%%%%%%%%%%%%%%%%%%%%%%%%%%%%%%%%%%%%%%%%%%%%%%%%%%%%%%%%%%%%%%%%%%%%%%%%%%%%%%%%%%%%%%%%%%%%%%%%%%%%%%%%%%%%%%%%
\section{Introduction}\label{Section:Introduction}
%%%%%%%%%%%%%%%%%%%%%%%%%%%%%%%%%%%%%%%%%%%%%%%%%%%%%%%%%%%%%%%%%%%%%%%%%%%%%%%%%%%%%%%%%%%%%%%%%%%%%%%%%%%%%%%%%%%%%%%%%%%%%%%
%%%%%%%%%%%%%%%%%%%%%%%%%%%%%%%%%%%%%%%%%%%%%%%%%%%%%%%%%%%%%%%%%%%%%%%%%%%%%%%%%%%%%%%%%%%%%%%%%%%%%%%%%%%%%%%%%%%%%%%%%%%%%%%

Developments in quantum technologies are increasingly important nowadays for various applications in sensing, transmission and processing of information. However, protection of quantum systems from unwanted interactions with the environment remains a major challenge. Dynamical decoupling (DD) is a widely used approach that aims to compensate the unwanted qubit-environment coupling by applying continuous fields or sequences of pulses \cite{Viola09PRL,Suter16RMP}.

Continuous dynamical decoupling (CDD), where the system is driven with a protecting dressing field for the entire duration of the experiment has already been demonstrated to compensate for noise sources in various media, e.g., in color centres in diamond and trapped ions \cite{Suter16RMP,WrachrupNano2013,DegenARPC2014,DegenRMP2017,BalasubramanianNatMat2009,deLangeScience2010,WrachtrupPRB2011,
KnowlesNatMat2014,HollenbergNatNano2011,WalsworthNature2013,LukinNature2013,BalasubramanianOpinBio2014,HirosePRA2012,AielloNatComm2013,StarkNatComm2017,StarkSciRep2018}. Then, an energy gap is opened in the dressed state basis, which allows for first order protection against noise in a perpendicular
direction of the applied field. However, the energy gap in the dressed state basis suffers from driving field fluctuations, resulting in additional noise which is not
compensated for. One way to overcome this problem is by applying
concatenated CDD, in
which one uses multiple additional (higher order) dressing fields to open smaller perpendicular energy gaps iteratively to compensate for the driving field noise of the previous order dressing field \cite{CaiNJP2012}.
As the driving field noise is usually proportional to the amplitude of the applied field, the application of additional dressing fields with their lower amplitudes leads to longer coherence times.
Major drawbacks of this approach are the complexity
due to operation with multiple fields and the lower energy gap in the highest order dressed basis, which leads to slower operation.
% as all quantum operators must be slower than it.
Apart from dressing field concatenation, an alternative approach was proposed recently, where a time dependent phase was added to a continuous driving field, yielding a time-dependent
detuning \cite{CohenFP2017}.
% The latter acts like the second driving field of the concatenated CDD but its noise can be negligible in some experimental implementations, e.g., in the microwave or radio-frequency regime when we can apply an arbirtrary waveform generator.
The latter acts like the second driving field of the concatenated CDD and its noise can be negligible in some experiments at the expense of complexity of implementation, e.g., of the time-dependent detuning. Other robust CDD schemes are also available when we consider multi-level systems \cite{StarkSciRep2018,CohenFP2017,TimoneyNature2011,AharonPRL2013,AharonNJP2016,BarfussNPhys2018}.

Dynamical decoupling by sequences of time-separated pulses (pulsed DD) is another widely used approach for compensation of environmental noise \cite{Suter16RMP}. Some disadvantages of pulsed DD include the requirements for both high pulse powers to limit dephasing during a pulse, and high repetition rate, so that refocusing is much faster than the correlation time of the environment \cite{Suter16RMP,CaiNJP2012}. Another important challenge are pulse imperfections, e.g., due to inhomogeneous broadening, low bandwidth, or field errors.
%Their detrimental impact is enhanced by the very large number of pulses and it often exceeds the effect of the environment \cite{RDD_review12Suter}.
One way to overcome these challenges is the application of sequences of pulses, where the error of one pulse can be compensated by the other pulses in a sequence by suitably choosing their relative phases \cite{RDD_review12Suter}, even to an arbitrary order \cite{GenovPRL2017,Sriarunothai2019QST}. The error compensation mechanism is usually based on composite pulses \cite{Levitt97Review,Tycko84,Schraft13PRA,Heinze13PRL,Torosov11PRA,Torosov11PRL,Alvarez13PRA,Genov2014PRL,Genov2018PRA}, which typically require that a sequence duration is much shorter than the correlation time of the environment for its self-compensatory mechanisms to work.
%These were also successfully demonstrated for DD for optical data storage in doped solids, where they outperformed the state-of-the art robust DD sequences \cite{GenovPRL2017}.
%

In this work, we propose a mixed approach between continuous and pulsed DD, i.e., mixed DD (MDD), where we use a strong continuous driving field to compensate a relatively fast noise in a perpendicular direction to the field and apply robust sequences of phased pulses in the dressed basis to protect against field noise. The latter refocus both the (usually slower) continuous driving field noise and systematic errors due to the environment.
We demonstrate a superior performance
of MDD with the same (or lower) average power of the driving fields
in comparison to standard concatenated CDD
for realistic experimental conditions in NV centers. The improvement is present even when the second driving field of concatenated CDD is assumed to be noiseless, which corresponds to the time dependent detuning scheme \cite{CohenFP2017}. Finally, we also demonstrate how our technique can be utilized for improved sensing.

MDD can be especially useful for systems where the correlation time of the environment is too short to apply robust sequences of pulses directly in the bare basis. Then the first strong (noisy) driving field suppresses the fast noise of the environment. This allows one to use longer robust DD sequences in the dressed basis to compensate for the relatively slower noise of the driving field.
MDD also allows the first driving field (and its noise) to be stronger in comparison to double-drive concatenated CDD as the robust DD sequences have a wider error compensation range than a simple second continuous drive.

The paper is organized as follows: in section \ref{Section:CCDD} we give a brief introduction to the idea of concatenated CDD. Then, we introduce the concept of MDD and compare its performance to concatenated CDD in section \ref{Section:MDD}. Finally, we propose how our technique can be utilized for improved sensing in the pulsed and continuous regime
in section \ref{Section:Sensing}, which is followed by a discussion.

%%%%%%%%%%%%%%%%%%%%%%%%%%%%%%%%%%%%%%%%%%%%%%%%%%%%%%%%%%%%%%%%%%%%%%%%%%%%%%%%%%%%%%%%%%%%%%%%%%%%%%%%%%%%%%%%%%%%%%%%%%%%%%%
%%%%%%%%%%%%%%%%%%%%%%%%%%%%%%%%%%%%%%%%%%%%%%%%%%%%%%%%%%%%%%%%%%%%%%%%%%%%%%%%%%%%%%%%%%%%%%%%%%%%%%%%%%%%%%%%%%%%%%%%%%%%%%%
\section{Concatenated Continuous Dynamical Decoupling}\label{Section:CCDD}
%%%%%%%%%%%%%%%%%%%%%%%%%%%%%%%%%%%%%%%%%%%%%%%%%%%%%%%%%%%%%%%%%%%%%%%%%%%%%%%%%%%%%%%%%%%%%%%%%%%%%%%%%%%%%%%%%%%%%%%%%%%%%%%
%%%%%%%%%%%%%%%%%%%%%%%%%%%%%%%%%%%%%%%%%%%%%%%%%%%%%%%%%%%%%%%%%%%%%%%%%%%%%%%%%%%%%%%%%%%%%%%%%%%%%%%%%%%%%%%%%%%%%%%%%%%%%%%
%

%%%%%%%%%%%%%%%%%%%%%%%%%% FIGURE 1 %%%%%%%%%%%%%%%%%%%%%%%%%%%%%
\begin{figure}[t!]
\includegraphics[width=\columnwidth]{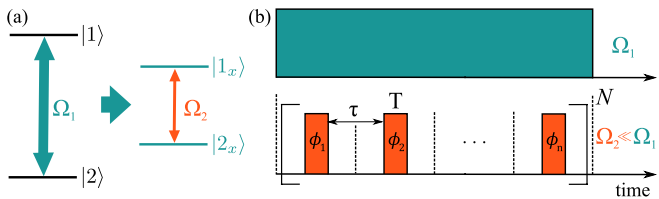}
\caption{(color online)
(a) Scheme for MDD implementation. The first driving field $\Omega_1$ is applied continuously and creates the first order dressed states $|1_{x}\rangle$ and $|2_{x}\rangle$, which are protected (to first order) from the environmental noise. However, they suffer from noise in $\Omega_1$. The robust DD sequence of phased pulses in the dressed basis makes use of the (usually) longer correlation time of the field noise and protects from the field noise in $\Omega_1$ and $\Omega_2$. (b) Scheme for experimental implementation with the switching of the two fields synchronized. The optimum ratio between the two peak Rabi frequencies depends on the particular experiment and is currently taken to be $\Omega_2/\Omega_1=0.1$, similarly to standard concatenated DD experiments in NV centers. The pulse separation time $\tau$ can also be zero.}
\label{Fig1:MDD_scheme}
\end{figure}
%%%%%%%%%%%%%%%%%%%%%%%%%% FIGURE 1 %%%%%%%%%%%%%%%%%%%%%%%%%%%%%

We consider a two-state system with a noise in the $z$ direction and a free evolution Hamiltonian ($\hbar=1$)
\begin{equation}
H_0=\frac{\sigma_{z}}{2}\left(\omega_0+\delta(t)\right),
\end{equation}
where $\omega_0$ is the Bohr transition frequency and the time-dependent detuning $\delta(t)$ is due to the noise and causes dephasing. We note that $\delta(t=0)$ can also be non-zero, i.e., we can have inhomogeneity of the resonance frequency, e.g., due to inhomogeneous broadening. In order to protect against dephasing due to the environment, we apply a resonant, noisy driving field in a perpendicular direction of the noise \cite{CaiNJP2012}
\begin{equation}
H_1=\Omega_1 (1+\epsilon_1(t))\cos{(\omega_0 t)}\sigma_{x},
\end{equation}
where $\Omega_1$ is the Rabi frequency of the driving field and $\epsilon_1(t)$ is the relative error due to field noise. We note that $\epsilon_1(t=0)$ can also be non-zero, i.e., we can have inhomogeneity of the applied field.

Usually, it is more straightfoward to consider the system in the interaction basis, rotating at $\omega_0$, i.e., an effective Hamiltonian with respect to $H_0^{(1)}=\omega_0 \sigma_{z}/2$, and apply the rotating-wave approximation ($\Omega_1\ll\omega_0$). Then, the evolution of the system with a Hamiltonian $H_0+H_1$ in the bare basis is described in the interaction basis by
\begin{equation}
H_{I1}=\frac{1}{2}\left[\delta(t)\sigma_{z}+\Omega_1 (1+\epsilon_1(t))\sigma_{x}\right],
\end{equation}
In the ideal case of a perfect driving field, the energy gap in the corresponding dressed basis due to the strong Rabi frequency $\Omega_1$ in effect suppresses the effect of the noise $\delta(t)$ \cite{CaiNJP2012}. However, in real experimental situations the driving field noise $\epsilon_1(t)$ is non-zero and itself causes dephasing.

In order to compensate for this noise, a second continuous driving field can be applied, as proposed in \cite{CaiNJP2012}. The Hamiltonian of this second driving field is given in the bare basis as:
\begin{equation}
H_2=2\Omega_2 (1+\epsilon_2(t))\cos{\left(\omega_0 t+\frac{\pi}{2}\right)}\cos{(\Omega_1 t)}\sigma_{x},
\end{equation}
where $\Omega_2$ is the Rabi frequency of the second driving field and $\epsilon_2(t)$ characterizes its noise. Again, we note that $\epsilon_2(t=0)$ can be non-zero, i.e., we can have inhomogeneity of the applied second field.
% We also assume in the following analysis that the noise realizations of $\epsilon_1(t)$ and $\epsilon_2(t)$ are independent although their noise characteristics are the same.
Then, the evolution of the system with a Hamiltonian $H_0+H_1+H_2$ in the bare basis is described in the interaction basis with respect to $H_0^{(1)}$ by
\begin{equation}
\widetilde{H}_{I1}=H_{I1}+\Omega_2 (1+\epsilon_2(t))\sigma_{y}\cos{(\Omega_1 t)}.
\end{equation}

%%%%%%%%%%%%%%%%%%%%%%%%%% FIGURE 2 %%%%%%%%%%%%%%%%%%%%%%%%%%%%%
\begin{figure*}[t]
\includegraphics[width=\textwidth]{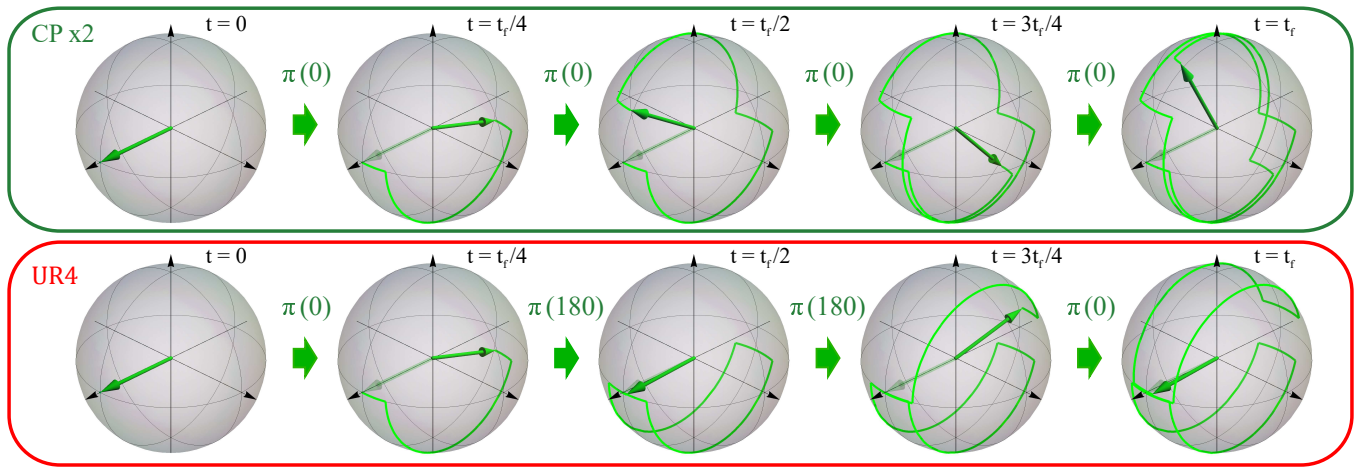}
\caption{(color online)
Bloch sphere representation of coherent evolution of the quantum state of an atom in the second order interaction basis as a result of a DD sequence of four time-separated phased pulses for (top) the Carr-Purcell (CP) sequence of two pulses with phases $\phi_{k}=0$, repeated twice, and (bottom) a UR4 sequence with phases $\phi_1=\phi_4=0$, $\phi_2=\phi_3=180^{\circ}$, as defined in \cite{GenovPRL2017}. The green vector shows the Bloch vector during the process at different times. The error adds up after every pulse for CP, while the specific evolution path of UR4 allows for a highly robust performance. The simulation is based on the Hamiltonian in Eq. \eqref{Eq:HI2} and
rectangular DD pulses with a Rabi frequency of $\Omega_2=2\pi~0.5$ MHz, $\epsilon_1 \Omega_1=0.1~\Omega_2$, $\epsilon_2=-0.1$, pulse duration $T=0.5~\mu$s, pulse separation $\tau= 3~\mu$s, and a storage time of $t_{f}=28~\mu$s. We assumed that $\epsilon_{i}=\text{const}$ in this simulation.}
\label{Fig2:CP_UR4}
\end{figure*}
%%%%%%%%%%%%%%%%%%%%%%%%%% FIGURE 2 %%%%%%%%%%%%%%%%%%%%%%%%%%%%%

Then, we consider the system in the second order interaction basis with respect to $H_0^{(2)}=\Omega_1 \sigma_{x}/2$, after applying the RWA ($\Omega_2\ll\Omega_1$) and assuming that the effect of the environment noise $\delta(t)$ can be neglected due to the first strong driving field. Thus, the effective Hamiltonian becomes \cite{CaiNJP2012}
\begin{equation}\label{Eq:HI2}
\widetilde{H}_{I2}=\frac{\Omega_1}{2}\epsilon_1(t)\sigma_{x}+\frac{\Omega_2}{2}(1+\epsilon_2(t))\sigma_{y},
\end{equation}
where $\Omega_2$ is the Rabi frequency of the second driving field and $\epsilon_2(t)$ is the relative error due to the second field noise. In the ideal case where $|\epsilon_1(t)\Omega_1|\ll\Omega_2$ and $\epsilon_2(t)\to 0$, the Hamiltonian takes the form $\widetilde{H}_{I2}\approx\Omega_2\sigma_{y}/2$, i.e., the effect of the environment noise and field noise of the first field are both suppressed by the concatenated CDD.

In a real experiment, however, $\epsilon_2(t)$ cannot be neglected and $|\epsilon_1(t)\Omega_1|\ll\Omega_2$ might not be feasible especially as the RWA approximation requires $\Omega_2\ll\Omega_1$. Then, the efficiency of concatenated CDD is reduced. One approach to counter the first limitation is to apply additional driving fields with the resulting increasing complexity of implementation \cite{CaiNJP2012}. Another alternative is to minimize the second order field noise $\epsilon_2(t)$ by adding a time-dependent detuning of the first field \cite{CohenFP2017}. In addition, one can use a stronger second drive with $\Omega_2\approx\Omega_1$ by changing its frequency to account for the Bloch-Siegert shift when RWA is not applicable \cite{AharonArxiv2018}. Next, we propose an alternative approach by applying phased pulsed sequences in the first order interaction basis, which compensate both field noise and environmental noise and do not require additional driving fields.

%%%%%%%%%%%%%%%%%%%%%%%%%%%%%%%%%%%%%%%%%%%%%%%%%%%%%%%%%%%%%%%%%%%%%%%%%%%%%%%%%%%%%%%%%%%%%%%%%%%%%%%%%%%%%%%%%%%%%%%%%%%%%%%
%%%%%%%%%%%%%%%%%%%%%%%%%%%%%%%%%%%%%%%%%%%%%%%%%%%%%%%%%%%%%%%%%%%%%%%%%%%%%%%%%%%%%%%%%%%%%%%%%%%%%%%%%%%%%%%%%%%%%%%%%%%%%%%
\section{Mixed Dynamical Decoupling}\label{Section:MDD}
%%%%%%%%%%%%%%%%%%%%%%%%%%%%%%%%%%%%%%%%%%%%%%%%%%%%%%%%%%%%%%%%%%%%%%%%%%%%%%%%%%%%%%%%%%%%%%%%%%%%%%%%%%%%%%%%%%%%%%%%%%%%%%%
%%%%%%%%%%%%%%%%%%%%%%%%%%%%%%%%%%%%%%%%%%%%%%%%%%%%%%%%%%%%%%%%%%%%%%%%%%%%%%%%%%%%%%%%%%%%%%%%%%%%%%%%%%%%%%%%%%%%%%%%%%%%%%%

The main idea of MDD is to combine continuous DD with a strong (noisy) field and a robust sequence of phased pulses to protect from both field and environmental noise.
Specifically, we apply the same first driving field as with standard CDD, but we replace the second continuous driving field with a robust sequence of phased pulses (see Fig. \ref{Fig1:MDD_scheme}). Unlike standard pulsed DD, the pulses are applied to perform flips in the second order interaction basis, as defined in the previous section. Then, the phases of the individual pulses are used to compensate both the usually slower noise of the first continuous field and make performance robust to noise from the second field. The main advantage in comparison to standard pulsed DD where the phased pulses are applied in the bare basis to compensate directly for environmental noise is the usually longer correlation time of the field noise, which allows for the application of long, highly robust sequences and more precise control of the relative phases between the pulses.

%%%%%%%%%%%%%%%%%%%%%%%%%% FIGURE 4 %%%%%%%%%%%%%%%%%%%%%%%%%%%%%
\begin{figure*}[t!]
\includegraphics[width=\textwidth]{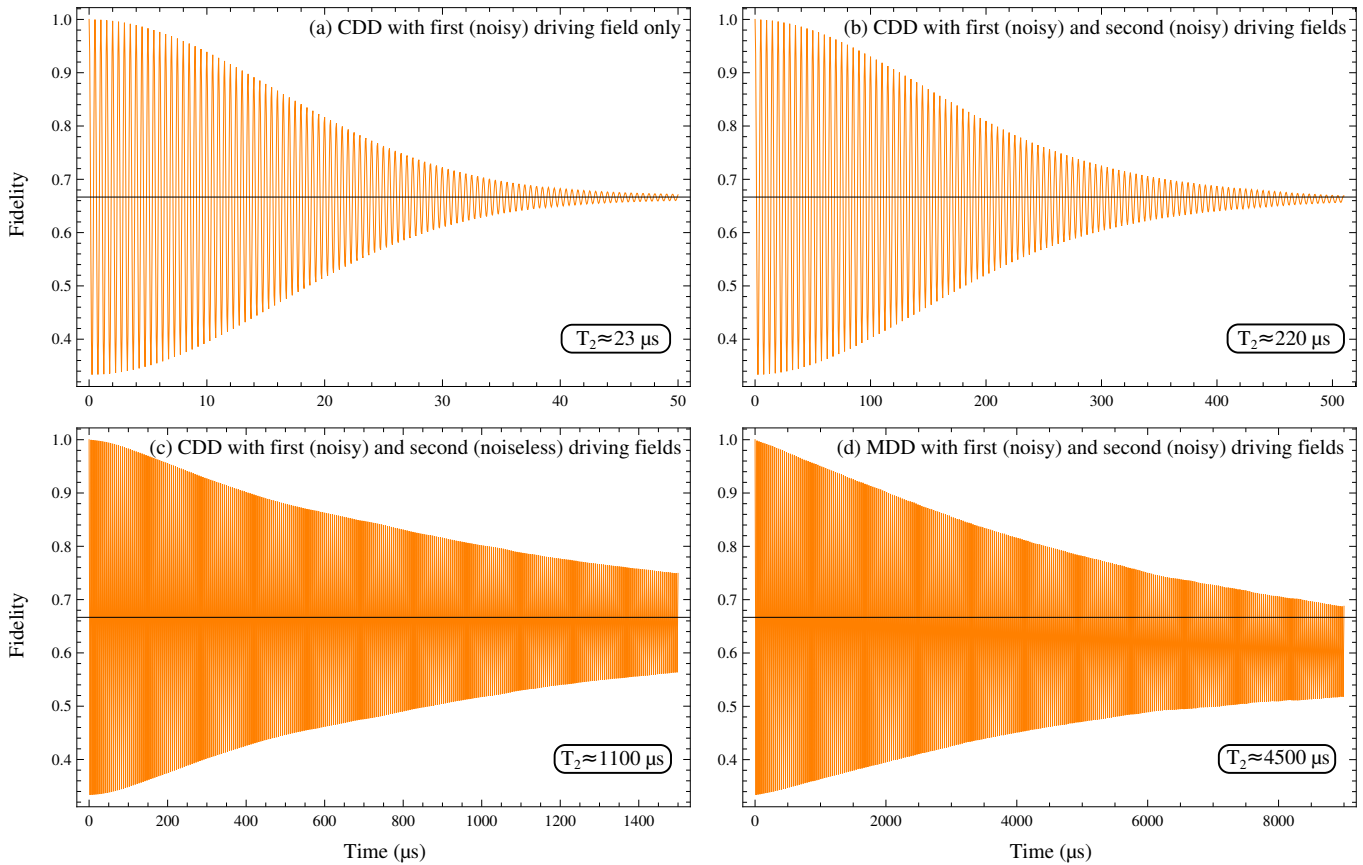}
\caption{(color online)
Simulation of DD performance for different DD sequences: (a) CDD with a single noisy field with $\Omega_1=2\pi~2$MHz, (b) Concatenated CDD with two noisy fields with $\Omega_1=2\pi~2$ MHz, $\Omega_2/\Omega_1=0.1$ , (c) Concatenated CDD with two fields with $\Omega_1=2\pi~2$ MHz, $\Omega_2/\Omega_1=0.1$, where the first field is assumed noisy and the second one - ideal, (d) Mixed DD with two noisy fields with $\Omega_1=2\pi~2$ MHz, $\Omega_2/\Omega_1=0.1$ and UR10 pulsed DD sequence of the second field with pulse duration of $T=2.5 \mu$s and pulse separation of $\tau=0.5 \mu$s. The UR10 sequence consists of ten pulses with phases $(0,4,2,4,0,0,4,2,4,0)\pi/5$ \cite{GenovPRL2017}, repeated during the storage time. The slight asymmetry of the MDD profile with respect to the quantum limit is due to the resolution of the simulation.}
\label{Fig4:DD_seqs}
\end{figure*}
%%%%%%%%%%%%%%%%%%%%%%%%%% FIGURE 4 %%%%%%%%%%%%%%%%%%%%%%%%%%%%%

As we apply the same first driving field as with standard CDD, the Hamiltonians $H_0$ and $H_1$ remain the same as in the previous section, while the Hamiltonian due to a single (phased) pulse in the dressed basis takes the form:
\begin{equation}\label{Eq:H2_definition}
H_2(\phi)=2\Omega_2(1+\epsilon_2(t))\cos{\left(\omega_0 t+\frac{\pi}{2}\right)}\cos{(\Omega_1 t + \phi)}\sigma_{x},
\end{equation}
i.e., we have added a phase parameter $\phi$. We note that the pulses in the dressed basis need not be rectangular or time-separated. The only requirement is a very good control over the (discrete) phase $\phi$, which is usually experimentally feasible.

The effective Hamiltonian in the second order interaction basis with respect to $H_0^{(2)}=\Omega_1 \sigma_{x}/2$, after applying the RWA ($\Omega_2\ll\Omega_1$) and assuming that the effect of the environment noise $\delta(t)$ can be neglected due to the first strong driving field, is then
\begin{equation}\label{Eq:HI2phi}
\widetilde{H}_{I2}(\phi)=\frac{\Omega_1}{2}\epsilon_1(t)\sigma_{x}+\frac{\Omega_2}{2}(1+\epsilon_2(t))\left(\cos{\phi}\sigma_{y}+\sin{\phi}\sigma_{z}\right).
\end{equation}
The phase $\phi$ can be used as a control parameter for implementation of robust DD sequences. The latter are a widely applicable method for both pulse error and environmental noise suppression \cite{RDD_review12Suter}. The main idea is to choose the relative phases of the pulses in a suitable way, so that experimental errors due to a single pulse are compensated by the other pulses in a sequence \cite{RDD_review12Suter}, which can be achieved even to an arbitrary order \cite{GenovPRL2017}.

The robust sequences are often based on composite pulses, which are derived to compensate systematic errors for a static environment, i.e., infinite noise correlation time \cite{GenovPRL2017}. Thus, their self-compensatory mechanism usually works efficiently if the sequence duration is shorter than the correlation time of the noise \cite{RDD_review12Suter,GenovPRL2017}. As a result, there is a trade-off between longer sequences that compensate systematic errors better but suffer when their duration becomes comparable to the noise correlation time \cite{GenovPRL2017}.

One can obtain intuition about the error-compensatory mechanism in the approximation of static amplitude noise, i.e., when $\epsilon_{k}(t)=\epsilon_{k},~k=1,2$ is constant. Then, double-drive CDD exhibits errors due to the noise of the second drive and also remains susceptible to a second order noise term $\sim \epsilon_1^2\Omega_1^2/\Omega_2$ even when the second field is noiseless (see Appendix, sec. \ref{Section:Fidelity_errors}). Both errors can be reduced to an arbitrarily high order by applying robust sequences of phased pulses in the dressed basis as long as the correlation time of the amplitude noise is long enough, which is usually feasible.
Figure \ref{Fig2:CP_UR4} shows an example of a comparison between the performance of two DD sequences: the Carr-Purcell (CP) and UR4, as described in \cite{CPMG_papers,GenovPRL2017}. It is evident that the distance between the initial and final Bloch vectors adds up after every pulse for CP, while the specific evolution path of UR4 allows for an almost perfect performance. A detailed analytic comparison of CP and UR4 in the approximation of a constant environment is given in the Appendix, sec. \ref{Section:Fidelity_errors}.

We note that some DD sequences improve performance only for specific initial states \cite{Suter16RMP,CPMG_papers,Genov2018PRA}. For example, the CP sequence in Fig. \ref{Fig2:CP_UR4} works very well for initial states aligned along the $y$ axis of the Bloch sphere and is then termed CPMG \cite{Suter16RMP,CPMG_papers,Genov2018PRA}. %Then, the Hamiltonian $\widetilde{H}_{I2}$ in Eq. \eqref{Eq:HI2phi} commutes with the initial state of the system.
As another example, if our system is initially in state $\sigma_{x}$,
i. e., its initial density matrix is $\rho_{I2}(t_{\text{i}})=\rho_{x}\equiv(\sigma_0+\sigma_{x})/2$),
it will not be affected by $\epsilon_1(t)$ noise even if we do not apply a second field as $\partial \rho_{I2}(t)=-\i[\rho_{I2}(t),\widetilde{H}_{I2}(\Omega_2=0)]=0$.
In order to %avoid these limitations and
ensure fair comparison of DD performance, we will use a measure of the fidelity, which does not depend on the initial state of the system and is based on \cite{BowdreyPhLett2002}:
\begin{equation}\label{Eq:Fid_definition}
F=\frac{1}{3}\sum_{k=x,y,z} \text{Tr} \left( U_{I2}^{(n)}\rho_{k}\left(U_{I2}^{(n)}\right)^\dagger\rho_{k} \right),
\end{equation}
where the initial density matrices $\rho_{k}\equiv(\sigma_0+\sigma_{k})/2, k=1,2,3$, i.e., the fidelity is obtained by averaging the fidelities for the three initial states, corresponding to the axial pure states of the system.

We compare the performance of MDD and CDD by a numerical simulation for DD in a two-state system, subject to magnetic noise ($\delta(t)$) and uncorrelated power fluctuations of the driving fields (see Appendix, Sec. \ref{Section:Numerics} for more details). The parameters of the noise have the characteristics for typical experiments in NV centers, as described in \cite{CaiNJP2012,AharonNJP2016}. We show that our simulation agrees well with previous numerical results for NV centers \cite{AharonNJP2016} and exhibits the expected dephasing time of $T_{2}^{\ast}\approx 3 \mu$s (see Appendix, Sec. \ref{Section:Numerics}).

Next, we compare the performance of our MDD sequence with traditional sequences for CDD and concatenated CDD (see Fig. \ref{Fig4:DD_seqs}). The simulations show that applying CDD with a single driving field increases significantly the coherence time from $3~\mu$s to tens of microseconds (see Fig. \ref{Fig4:DD_seqs}a). The remaining decay is due to the noise of the driving field itself, as shown in previous publications \cite{CohenFP2017}. We note that we define the $T_2$ of the fidelity as the time it takes to drop to from $1$ to $\approx 0.79$, which corresponds to a $1/\e$ drop in the difference from the quantum limit of $0.67$. Applying a second driving field (concatenated CDD) expectedly leads to a substantial increase of the coherence time to approximately $220~\mu$s (see Fig. \ref{Fig4:DD_seqs}b). One can see from Fig. \ref{Fig4:DD_seqs}c that the noise in the second driving field is the main reason for the remaining decay, as assuming a noiseless second driving field boosts the coherence time even further to more than $1100~\mu$s. %Our simulations show that the MDD sequence significantly outperforms concatenated CDD even when the second field is assume to be noiseless and can preserve the fidelity of the quantum states for coherence times of the order of $4.5$ ms.

Next, we show that the MDD protocol, applied with a UR10 pulsed sequence in the first-order dressed basis, significantly outperforms concatenated CDD and achieves coherence times of the order of $4.5$ ms (see Fig. \ref{Fig4:DD_seqs}d). Thus, the simulations show that MDD boosts storage duration by more than $20$ times in comparison to standard double-drive concatenated CDD
for the same peak powers of the driving fields and even slightly lower average power for MDD due to the pulse separation of the second driving field
(compare with Fig. \ref{Fig4:DD_seqs}b). Additionally, MDD outperforms the concatenated CDD sequence even when we assume that the second driving field is noiseless (compare with Fig. \ref{Fig4:DD_seqs}c). The reason for the improved performance is that MDD compensates
(most of) the amplitude noise from both driving fields and much of the remnant environmental (magnetic) noise, which is not suppressed by the first driving field. For example, CDD with a noiseless second drive remains susceptible to a second order noise term of the first drive $\sim \epsilon_1^2\Omega_1^2/\Omega_2$, which MDD can in principle compensate to an arbitrary order.
More details on the UR10 sequence and its error-compensatory mechanism are described in the Appendix, sec. \ref{Section:Fidelity_errors} and \cite{GenovPRL2017}. We note that other robust phased sequences can also be applied, e.g., the widely used XY4, XY8 or KDD sequences \cite{Suter16RMP,DegenRMP2017}, and the optimum sequence will depend on the specific environment.

We note that the population relaxation time of an NV center can reach
up to $6$ ms \cite{Bar-GillNatComm2013},
so the fidelity at long storage times with MDD might be affected by the population relaxation time of the system  (not taken into account in the simulations) in this particular implementation.

%%%%%%%%%%%%%%%%%%%%%%%%%%%%%%%%%%%%%%%%%%%%%%%%%%%%%%%%%%%%%%%%%%%%%%%%%%%%%%%%%%%%%%%%%%%%%%%%%%%%%%%%%%%%%%%%%%%%%%%%%%%%%%%
%%%%%%%%%%%%%%%%%%%%%%%%%%%%%%%%%%%%%%%%%%%%%%%%%%%%%%%%%%%%%%%%%%%%%%%%%%%%%%%%%%%%%%%%%%%%%%%%%%%%%%%%%%%%%%%%%%%%%%%%%%%%%%%
\section{Sensing with MDD}\label{Section:Sensing}
%%%%%%%%%%%%%%%%%%%%%%%%%%%%%%%%%%%%%%%%%%%%%%%%%%%%%%%%%%%%%%%%%%%%%%%%%%%%%%%%%%%%%%%%%%%%%%%%%%%%%%%%%%%%%%%%%%%%%%%%%%%%%%%
%%%%%%%%%%%%%%%%%%%%%%%%%%%%%%%%%%%%%%%%%%%%%%%%%%%%%%%%%%%%%%%%%%%%%%%%%%%%%%%%%%%%%%%%%%%%%%%%%%%%%%%%%%%%%%%%%%%%%%%%%%%%%%%

Magnetometry experiments require the measurement of a signal whose amplitude is related to a magnetic field to be sensed. We demonstrate two approaches for utilizing MDD for sensing of an oscillating (AC) field in this section. Pulsed DD and concatenated CDD have already been applied for sensing AC fields \cite{DegenRMP2017,HirosePRA2012,AielloNatComm2013,StarkNatComm2017,StarkSciRep2018}. The application of MDD leads to an increase in the coherence time, which allows for higher sensitivity of the sensing protocol.

%%%%%%%%%%%%%%%%%%%%%%%%%% FIGURE 4 %%%%%%%%%%%%%%%%%%%%%%%%%%%%%
\begin{figure*}[t!]
\includegraphics[width=0.95\textwidth]{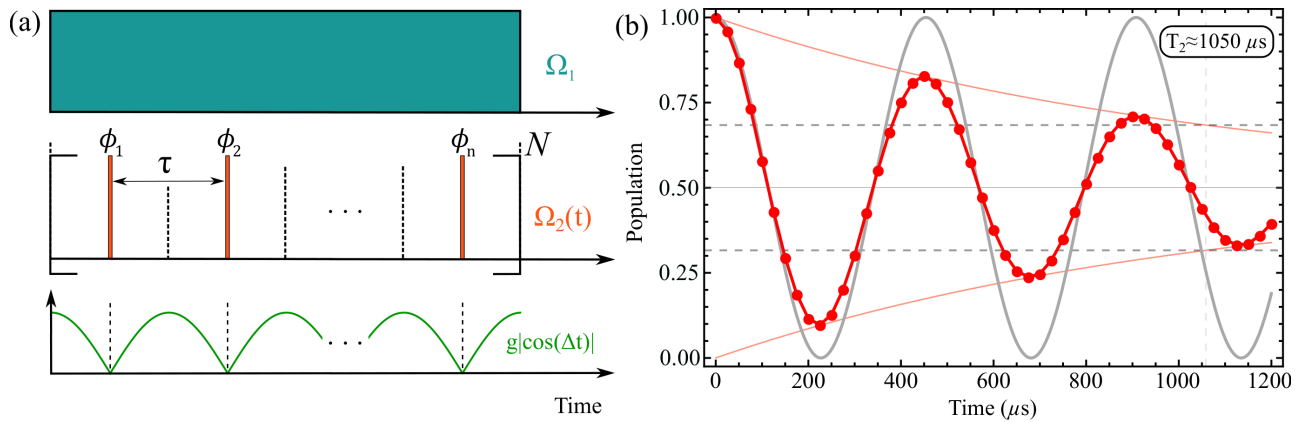}
\caption{(color online)
(a) Scheme for sensing with pulsed MDD where we detect the sensed field amplitude by measuring its effect during the free evolution time between the pulses. As the pulses are short, the effect of the sensed field when the second field is on is small and can be neglected. The DD pulses have a duration $T=\pi/\Omega_2$ and time separation of $\tau$, such that $\Delta(\tau+T)=\pi$. The accumulated pulse area in the toggling basis is $\Theta(t)=\int_0^t g|\cos{(\Delta t)}|$ and leads to Rabi oscillations, which can be observed directly in the bare basis stroboscopically at times, such that $\Omega_{1} t=0~(\text{mod}~2\pi)$ and $\int_{0}^{t}|\Omega_{2}(t^{\prime})|\d t^{\prime}=0 ~(\text{mod}~2\pi)$. (b) Simulation of MDD for sensing, showing the population of the ground state in the bare basis, taken stroboscopically at intervals of $25$ $\mu$s and corrected for expected population and phase evolution (see simulation details in the Appendix, Sec. \ref{Section:Numerics}). The frequency and amplitude noise is the same as in the simulations in Fig. \ref{Fig4:DD_seqs}. The first field is applied continuously with a Rabi frequency of $\Omega_1=2\pi~2$ MHz. The DD pulses follow the UR10 sequence \cite{GenovPRL2017} with $\Omega_2=2\pi~0.2$ MHz,  corresponding $\pi$-pulse duration $T=2.5$ $\mu$s and pulse separation of $22.5$ $\mu$s, such that $\tau+T=25 \mu$s. The sensed field has an amplitude $g=2\pi~6.92$ kHz, angular frequency shift $\Delta=2\pi~20$ kHz, and phase $\xi=0$. The light gray curve shows the respective theoretical evolution of the population $P=\cos(\Theta(t)/2)$, defined in Eq. \eqref{Theta_t}, for an ideal system without noise and with perfect instantaneous $\pi$ pulses. The red curve shows the evolution with non-instantaneous pulses and noise. %, obtained at times $m(T+\tau)$.
The coherence time is estimated $T_2 \approx 1050~\mu$s.
}
\label{Fig4pulsed}
\end{figure*}
%%%%%%%%%%%%%%%%%%%%%%%%%% FIGURE 4 %%%%%%%%%%%%%%%%%%%%%%%%%%%%%

First, we consider the Hamiltonian
\begin{align}
H_{\text{s}}=&\frac{\omega_0}{2}\sigma_{z}+\Omega_1\sigma_{x}\cos{(\omega_0 t)}+2\Omega_2(t)\sigma_{x}\cos{\left(\omega_0 t+\frac{\pi}{2}\right)}\notag\\
&\times\cos{(\Omega_1 t+\phi)}+g\sigma_{x}\cos{(\omega_{\text{s}}t+\xi)},
\end{align}
where $\omega_0$ is the Bohr transition frequency, $\Omega_1$ is the Rabi frequency of the first driving field and $\Omega_2$ is the (rescaled) Rabi frequency of the applied pulse with a phase $\phi$. This part of the Hamiltonian corresponds to the bare basis Hamiltonian $H_0+H_1+H_2$, as defined in Eq. \eqref{Eq:H2_definition}, where we have omitted here the noisy terms for compactness of notation. Additionally, $\omega_s=\omega_0+\Delta$ is the (angular) frequency of the signal, $g$ is its Rabi frequency, and $\xi$ is its unknown (random) phase. We then move to the interaction basis with respect to $H_0^{(1)}=\omega_0\sigma_{z}/2$ and obtain after applying the RWA ($\Omega_1\ll\omega_0$)
\begin{align}\label{H_sensing_1s}
\widetilde{H}_{\text{1,s}}=&\frac{\Omega_1}{2}\sigma_{x}+\Omega_2(t)\sigma_{y}\cos{(\Omega_1 t+\phi)}\\
&+\frac{g}{2}\left(\sigma_{x}\cos{(\Delta t+\xi)}+\sigma_{y}\sin{(\Delta t+\xi)}\right).\notag
\end{align}
We then move to the second order interaction basis with respect to $H_0^{(2)}=\Omega_1\sigma_{x}/2$ and obtain after applying the RWA ($\Omega_2\ll\Omega_1$) and neglecting the $\sigma_{y}$ term of the signal (we assume $\Delta\ll\Omega_1$ and $g\ll\Omega_1$)
\begin{equation}\label{H_sensing}
\widetilde{H}_{\text{2,s}}=\frac{\Omega_2}{2}\left(\sigma_{y}\cos{(\phi)}+\sigma_{z}\sin{(\phi)}\right)+\frac{g}{2}\sigma_{x}\cos{(\Delta t+\xi)}.
\end{equation}
%We note that the phase $\phi$ is also time-dependent but we assume that it changes instantaneously.

\begin{figure*}[t!]
\includegraphics[width=0.95\textwidth]{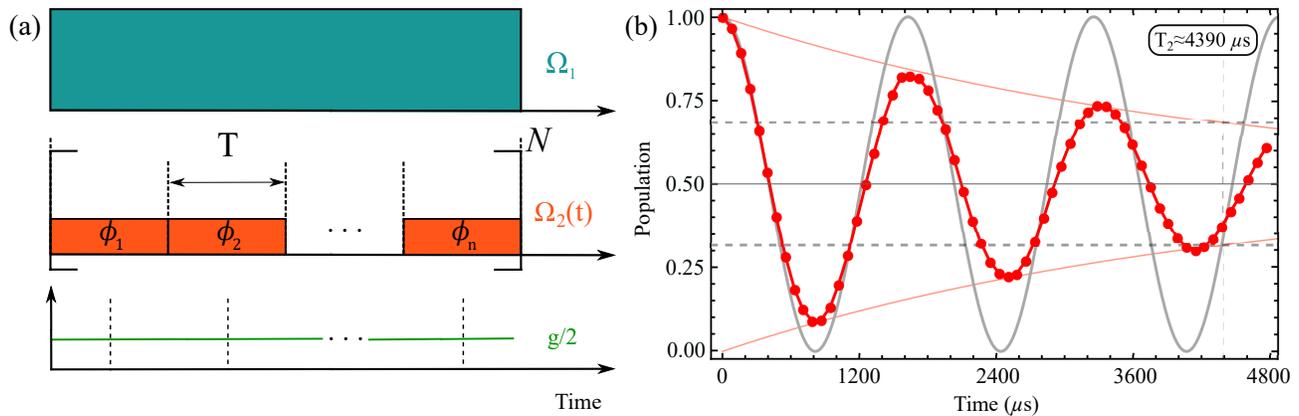}
\caption{(color online)
(a) Scheme for sensing with continuous phased field MDD.
%when the sensed field is detected only through its effect during the pulses.
The DD pulses have a duration $T=\pi/\Omega_2$ and zero separation ($\tau=0$). Similarly to pulsed MDD, we require $\Delta(\tau+T)=\pi$. The accumulated pulse area is $\Theta(t)=g t/2$ and leads to Rabi oscillations, which can be observed directly in the bare basis stroboscopically at times, such that $\Omega_{1} t=0~(\text{mod}~2\pi)$ and $\int_{0}^{t}|\Omega_{2}(t^{\prime})|\d t^{\prime}=0 ~(\text{mod}~2\pi)$. (b) Simulation of MDD for sensing, showing the population of the ground state in the bare basis taken stroboscopically at intervals of $80$ $\mu$s. The population is observed directly in the bare basis without corrections as the interval of $80$ $\mu$s corresponds to the duration of one UR10 sequence cycle (see simulation details in the Appendix, Sec. \ref{Section:Numerics}). The frequency and amplitude noise is the same as in the simulations in Fig. \ref{Fig4:DD_seqs}. The first field is applied continuously with a Rabi frequency of $\Omega_1=2\pi~2$ MHz. The DD pulses follow the UR10 sequence \cite{GenovPRL2017} with $\Omega_2=2\pi~62.5$ kHz, corresponding $\pi$-pulse duration $T=8$ $\mu$s. The sensed field has an amplitude $g=2\pi~2.46$ kHz, angular frequency shift $\Delta=\Omega_2$, and phase $\xi=0$. The light gray curve shows the respective theoretical evolution of $P=\cos(\Theta(t)/2)$, defined in Eq. \eqref{Theta_t}, in an ideal system without frequency and amplitude noise. The red curve shows the evolution with frequency and amplitude noise of both driving fields. The coherence time is estimated $T_2 \approx 4390~\mu$s.
}
\label{Fig5_UR10cw}
\end{figure*}

Next, we give two examples for sensing with MDD (see Figs. \ref{Fig4pulsed} and \ref{Fig5_UR10cw} for the respective experimental schemes). In both cases, we achieve improved sensing with MDD
 by choosing the duration $T$ of the DD pulses, each with (ideally) a pulse area of $\pi$, and their time separation $\tau$ to satisfy the condition
\begin{equation}\label{Eq:PMDD}
\Delta(\tau+T)=\pi,
\end{equation}
where $\Delta$ is in angular frequency units. In both cases, we prepare the system in
its ground state (in the bare basis). The effect of the sensed field leads to Rabi oscillations, which we can observe stroboscopically directly in the bare basis.

As a first example, we consider the case where we can neglect the effect of the sensed
field during the applied DD pulses in the dressed basis ($T\ll\tau$), and we require $\Delta\ll\Omega_2$ and $g\ll\Omega_2$. This is similar to pulsed DD applications without the first driving field. We can detect the sensed field by its effect between the pulses. Second, we consider the opposite regime ($T\gg\tau$) in the case of zero pulse separation, i.e.,  when we can replace the pulsed DD sequences in the dressed basis by a continuous phased field. Then, we require  $\Delta=\Omega_2$ and $g\ll\Omega_2$ and detect the sensed field by its effect during the pulses, i.e., during the interaction with the second phased field. It is in principle possible to apply the MDD scheme in the intermediate regime when the effect of the sensed field cannot be neglected both during and between the pulses, e.g., when $\tau=T$, but the effect of the sensed field is then more complicated.

%%%%%%%%%%%%%%%%%%%%%%%%%%%%%%%%%%%%%%%%%%%%%%%%%%%%%%%%%%%%%%%%%%%%%%%%%%%%%%%%%%%%%%%%%%%%%%%%%%%%%%%%%%%%%%%%%%%%%%%%%%%%%%%
\subsection{Pulsed MDD sensing}\label{Subsec:Pulsed_MDD}
%%%%%%%%%%%%%%%%%%%%%%%%%%%%%%%%%%%%%%%%%%%%%%%%%%%%%%%%%%%%%%%%%%%%%%%%%%%%%%%%%%%%%%%%%%%%%%%%%%%%%%%%%%%%%%%%%%%%%%%%%%%%%%%

First, we consider the case when $T\ll\tau$ and we can neglect the effect of the sensed
field during the applied DD pulses in the dressed basis. Thus, we detect the sensed field only by its effect during the free evolution time (in the dressed basis) between the pulses. In the approximation of instantaneous pulses $T\to 0$, the pulse separation is given by $\tau=\pi/\Delta$. The experimental scheme is shown in Fig. \ref{Fig4pulsed}(a).
One can acquire intuition about the sensing mechanism by considering a single period free evolution (for time $\tau/2$)-$\pi$ (phased) pulse-free evolution (for time $\tau/2$).
It proves useful to consider the toggling basis with respect to
%$H_0^{(3)}(t)=$
$\frac{\Omega_2(t)}{2}\left(\sigma_{y}\cos{(\phi)}+\sigma_{z}\sin{(\phi)}\right)$, where we have added a time dependence to $\Omega_2(t)$ to emphasize that $\Omega_2(t)=\Omega_2$ during a pulse and $\Omega_2(t)=0$ between the pulses.
The toggling basis unitary during the first period is just the identity operator,
\black
%$\widetilde{R}_{\text{tog}}(t<\tau/2)=I$,
so the Hamiltonian in the toggling frame before the $\pi$ pulse is given by
\begin{equation}\label{H_tog_1}
H_{\text{tog}}(t<\tau/2)=\widetilde{H}_{\text{2,s}}=\frac{g}{2}\sigma_{x}\cos{(\Delta t+\xi)}.
\end{equation}
We then applying a $\pi$ (phased) pulse and assume that effect of the sensed field during the pulse is negligible. Then, the Hamiltonian in the toggling frame after the $\pi$ pulse is given by
\begin{equation}\label{H_tog_2}
H_{\text{tog}}(t>\tau/2)=\widetilde{H}_{\text{2,s}}=-\frac{g}{2}\sigma_{x}\cos{(\Delta t+\xi)}.
\end{equation}
It is evident that the toggling frame Hamiltonian changes its sign due to the DD pulse and the effect of the sensed field can be cancelled, e.g., if $\Delta=0$. One way to avoid this cancellation is to require $\cos{(\Delta (-t-t_{\text{p,center}})+\xi)}=-\cos{(\Delta (t-t_{\text{p,center}})+\xi)}$, where $t_{\text{p,center}}$ is the center of each DD pulse (see also Fig. \ref{Fig4pulsed}). Thus, the DD pulses should be centered at the times when the function $\cos{(\Delta t+\xi)}$ changes its sign, which leads to the requirement in Eq. \eqref{Eq:PMDD}. Then, the toggling frame Hamiltonian during free evolution is
\begin{equation}\label{H_tog_free}
H_{\text{tog}}(t)=\frac{g}{2}\sigma_{x}|\cos{(\Delta t)}|.
\end{equation}

%%%%%%%%%%%%%%%%%%%%%%%%%% FIGURE 5 %%%%%%%%%%%%%%%%%%%%%%%%%%%%%
%%%%%%%%%%%%%%%%%%%%%%%%%% FIGURE 5 %%%%%%%%%%%%%%%%%%%%%%%%%%%%%

As a result of the signal, the sensing qubit will perform Rabi oscillations in the toggling basis.
We note that these correspond to phase accumulation of the $|\uparrow_{x}\rangle$ and $|\downarrow_{x}\rangle$ states due to the signal and Ramsey oscillations in the respective interaction basis, similarly to standard pulsed DD.
Assuming that the system is initially prepared in the ground state in the bare basis, the transition probability to the excited state in the toggling basis will depend on the pulse area $\Theta$ that is proportional to $g$ and takes the form
\begin{equation}\label{Theta_t}
P=\cos{(\Theta(t)/2)},~\Theta(t)\equiv\int_{0}^{t}g|\cos{(\Delta t^{\prime})}|\d t^{\prime}\approx\frac{2}{\pi}g t.
\end{equation}
We note that the Rabi oscillations in the toggling basis, observed stroboscopically at times when $\Omega_1 t=2\pi k,~k\in N$ correspond directly to Rabi oscillations in the $y-z$ plane in the bare basis up to well-defined shifts due to the number and phases of the pulses. As noise is present, it is best to measure at times $m(\tau+T), m\in N$ when it is (ideally) refocused. Finally, we also note that we assumed $\xi=0$ in the last two equations, which allows for maximum contrast. The scheme can also be applied for other $\xi$ but with a lower contrast, similarly to standard %Ramsey
experimental schemes for sensing with pulsed DD \cite{DegenRMP2017} or in combination with the Qdyne protocol \cite{Schmitt2017Science}.

%In summary, the sensing protocol proceeds as follows. Our goal is to measure the amplitude $g$ of an oscillating field with an angular frequency $\omega_{s}=\omega_0+\Delta$. First, we initialize the system in the ground state in the bare basis. Then, we apply a field with a Rabi frequency $\Omega_1$ and frequency $\omega_0$, synchronized with phased $\pi$-pulses of duration $T$ and pulse separation $\tau$ ($T\ll \tau$), such that $T+\tau=\pi/\Delta$. The sensed field causes Rabi oscillations in the toggling basis. The latter corresponds to rotation in the $y-z$ plane (population transfer) in the bare basis when observed stroboscopically at times when $t=2\pi k/\Omega_1, k\in N$. Due to noise, the times of the stroboscopic measurements should also satisfy $t=m(\tau+T), m\in N$ when the noise is (ideally) refocused. The population evolution in the toggling basis corresponds directly to population rotation with a pulse area $\Theta(t)$ in the bare basis up to well-defined population shifts, which depend on the number and phases of the $\pi$-pulses and can easily be calibrated.

Figure \ref{Fig4pulsed}(b) shows a simulation of evolution of the population of the ground state in the bare basis for sensing. The experimental parameters and all characteristics of frequency and amplitude noise are the same as in Sec. \ref{Section:MDD}. The only difference is the $\pi$-pulse separation, which is now taken to be equal to $\tau=9T$ in order to correspond to the experimental regime of the pulsed MDD where we assume $\tau\gg T$. We also note that the simulation results now show the population in the bare basis at times $m(\tau+T), m\in N$ when noise is (ideally) refocused. The results are corrected for the expected population changes and phase evolution due to the applied DD sequence. For example, at time $\tau+T=25~\mu$s, we have applied one $\pi$ pulse with the second driving field and the population in the bare basis will be inverted, which we correct for (see more details in the Appendix, Sec. \ref{Section:Numerics}). The simulation shows that the MDD scheme increases substantially the coherence time to $T_2\approx 1050~\mu$s, which is several times higher than with continuous double drive. We note that the longer pulse separation reduces slightly the coherence time in comparison to the MDD results in Sec. \ref{Section:MDD}. Finally, we note that we use now the standard definition of coherence time in sensing experiments, i.e., the time when the peak population drops to $P\approx 0.68$, which is $1/\e$ the difference from $1$ to the decoherence limit of equal population distribution. %We note that the slight shift in our estimate from expected evolution from theory is due to the non-instantaneous nature of our $\pi$ pulses. It is present even when no noise is assumed and can be taken into account in the actual estimation of $g$.

%%%%%%%%%%%%%%%%%%%%%%%%%%%%%%%%%%%%%%%%%%%%%%%%%%%%%%%%%%%%%%%%%%%%%%%%%%%%%%%%%%%%%%%%%%%%%%%%%%%%%%%%%%%%%%%%%%%%%%%%%%%%%%%
\subsection{Continuous field MDD sensing}\label{Subsec:Cont_MDD}
%%%%%%%%%%%%%%%%%%%%%%%%%%%%%%%%%%%%%%%%%%%%%%%%%%%%%%%%%%%%%%%%%%%%%%%%%%%%%%%%%%%%%%%%%%%%%%%%%%%%%%%%%%%%%%%%%%%%%%%%%%%%%%%

In this case, we assume that we cannot neglect the effect of the sensed field during the pulsed interaction. On the contrary, we will use this same effect for sensing. We apply a continuous field ($\tau=0$) with a Rabi frequency $\Omega_2$, where we change the phase of the field at equal intervals of time $T=\pi/\Omega_2$ (see Fig. \ref{Fig5_UR10cw}(a)). Additionally, we choose the detuning of the sensed field $\Delta=\Omega_2$.

First, we consider the Hamiltonian $\widetilde{H}_{\text{2,s}}$ in Eq. \eqref{H_sensing} during a time period when the phase $\phi$ is constant. We perform a rotation of our basis by $R_3(\phi)=\exp{(\i\phi\sigma_{x}/2)}$, so the Hamiltonian in the new rotated basis becomes
\begin{equation}\label{Eq:H_frame_3s}
\widetilde{H}_{\text{3,s}}=\frac{\Omega_2}{2}\sigma_{y}+\frac{g}{2}\sigma_{x}\cos{(\Omega_2 t+\xi)}.
\end{equation}
We then move to the interaction picture with respect to $H_0^{(4)}=\Omega_2\sigma_{y}/2$ by the rotation $R_4(t)=\exp{(\i\Omega_2 t\sigma_{y}/2)}$ and obtain the Hamiltonian
\begin{align}
\widetilde{H}_{\text{4,s}}&=\frac{g}{4}\left[\sigma_{x}\cos{(\xi)}-\sigma_{z}\sin{(\xi)}\right]\\
&+\frac{g}{4}\left[\sigma_{x}\cos{(\xi+2\Omega_2 t)}+\sigma_{z}\sin{(\xi+2\Omega_2 t)}\right].\notag
\end{align}
Then, we neglect the fast-rotating terms at angular frequency $2\Omega_2$ ($g\ll\Omega_2$), and the Hamiltonian in RWA becomes
\begin{equation}\label{Eq:H_frame_4s}
\widetilde{H}_{\text{4,s}}\approx\frac{g}{4}\left[\sigma_{x}\cos{(\xi)}-\sigma_{z}\sin{(\xi)}\right],
\end{equation}
The time evolution due to this Hamiltonian causes rotation around an axis at an angle $\xi$ (from the $x$ axis) in the $x$-$z$ plane of the Bloch sphere in this basis. If the phase $\phi$ is kept constant, this case corresponds to standard concatenated CDD (see Sec. \ref{Section:CCDD} and Fig. \ref{Fig4:DD_seqs}(b)) with the respective characteristics of environment noise suppression \cite{CaiNJP2012}. %If the phase $\phi$ remains constant during the whole interaction,
In this case the obtained contrast is independent from the initial phase $\xi$ of the sensed field. % as its variation only changes the rotation axis in the $x$-$z$ plane.
However, MDD requires phase changes for more robust noise suppression (see Sec. \ref{Section:MDD}), which makes contrast dependent on the initial phase $\xi$ of the signal (see Appendix \ref{Section:MDD_phase_selectivity} for a detailed discussion). The best results require that the signal $g(t)$ changes its sign at the center of every $\pi$ pulse (equivalently in the middle of every interval of constant phase evolution), which is exactly the same condition as with pulsed MDD. Thus, assuming that the phase changes occur at times $t_{m}=m T,~m\in N$, where $T=\pi/\Omega_2$ (see Fig. \ref{Fig5_UR10cw}(a)), the best contrast is obtained for $\xi=0$ and the Hamiltonian in Eq. \eqref{Eq:H_frame_4s} becomes
\begin{equation}\label{Eq:H_frame_4s_MDD}
\widetilde{H}_{\text{4,s}}^{\text{MDD}}=g\sigma_{x}/4.
\end{equation}
As a result, the sensing qubit will perform Rabi oscillations in the $y-z$ plane of the Bloch sphere in the basis of the Hamiltonian in Eq. \eqref{Eq:H_frame_4s}. Assuming that it is initially prepared in the ground state, the transition probability to the excited state will depend on the pulse area $\widetilde{\Theta}$ that is proportional to $g$ and takes the form
\begin{equation}\label{Theta_t}
P=\cos{(\widetilde{\Theta}/2)},~\widetilde{\Theta}\equiv g t/2.
\end{equation}
We note that the Rabi oscillations in the basis of the Hamiltonian in Eq. \eqref{Eq:H_frame_4s}, observed stroboscopically at times when $\Omega_1 t=2\pi k,~k\in N$ correspond directly to Rabi oscillations in the $y-z$ plane in the bare basis up to well-defined shifts due to the number and phases of the pulses. As noise is present, it is best to measure at times $m(\tau+T)=m T, m\in N$ when it is (ideally) refocused.

Finally, we note that we can choose the ``good'' initial phase $\xi$ of the signal by shifting the whole MDD sequence in time. We only need to change the duration of the first pulse period to satisfy $\xi+\Omega_2(t_1-t_0)=0~\text{mod}(\pi)$, where $t_0$ and $t_1$ are the initial and final times of the first period with a constant phase and apply the subsequent phase changes at intervals of $T$. This allows for the combination continuous MDD with other sensing techniques, e.g., the Qdyne protocol \cite{Schmitt2017Science}.

Figure \ref{Fig5_UR10cw}(b) shows a simulation of evolution of the population of the ground state in the bare basis for sensing. The experimental parameters and all characteristics of frequency and amplitude noise are the same as in Subsec. \ref{Subsec:Pulsed_MDD}. The only differences are the $\pi$-pulse separation, which is now taken to be equal to $\tau=0$, and the magnitude of the Rabi frequency of the second drive $\Omega_2$, which is reduced to ensure that the average power input of the continuous and pulsed MDD regimes are the same. We again use the standard definition of coherence time in sensing experiments, similarly to Subsec. \ref{Subsec:Pulsed_MDD}.

The simulation shows that the coherence time is increased substantially to $T_2\approx~4.4$ ms, which is much higher in comparison to both pulsed MDD and standard double-drive CDD. The zero pulse separation improves performance in comparison to pulsed MDD as the lack of free evolution reduces the effect of high frequency noise. Furthermore, the application of robust phased sequences allows to improve robustness even though the Rabi frequency of the second drive is only $\Omega_2\approx 0.03~\Omega_1$ and is 3.2 times lower than with pulsed MDD (to maintain the same average power input). Thus, the effect of amplitude noise in the first drive would be high without the phases. The numerical simulations show that decay is due to high order frequency noise $\delta$ and amplitude noise in $\Omega_1$. We also observe a small shift in the estimated amplitude of g, mainly due to high-order noise in $\Omega_1$, which might be possible to compensate with higher order robust sequences. Nevertheless, the coherence time of $T_2\approx~4.4$ ms is several times higher than the other schemes and approaches the population lifetime of an NV center. The latter
can reach up to $6$ ms \cite{Bar-GillNatComm2013} and is not taken into account in the simulation.

%%%%%%%%%%%%%%%%%%%%%%%%%%%%%%%%%%%%%%%%%%%%%%%%%%%%%%%%%%%%%%%%%%%%%%%%%%%%%%%%%%%%%%%%%%%%%%%%%%%%%%%%%%%%%%%%%%%%%%%%%%%%%%%
%%%%%%%%%%%%%%%%%%%%%%%%%%%%%%%%%%%%%%%%%%%%%%%%%%%%%%%%%%%%%%%%%%%%%%%%%%%%%%%%%%%%%%%%%%%%%%%%%%%%%%%%%%%%%%%%%%%%%%%%%%%%%%%
\section{Discussion}\label{Section:Discussion}
%%%%%%%%%%%%%%%%%%%%%%%%%%%%%%%%%%%%%%%%%%%%%%%%%%%%%%%%%%%%%%%%%%%%%%%%%%%%%%%%%%%%%%%%%%%%%%%%%%%%%%%%%%%%%%%%%%%%%%%%%%%%%%%
%%%%%%%%%%%%%%%%%%%%%%%%%%%%%%%%%%%%%%%%%%%%%%%%%%%%%%%%%%%%%%%%%%%%%%%%%%%%%%%%%%%%%%%%%%%%%%%%%%%%%%%%%%%%%%%%%%%%%%%%%%%%%%%

In our analysis, we showed that MDD can improve performance in comparison to other well known techniques for DD, e.g., double-drive concatenated CDD. As MDD combines continuous and robust phased sequences of pulses, where each has an effect of a DD noise filter, its overall positive effect depends on the effect of this
combined noise filter.

MDD can be particularly useful in case of high frequency environmental (magnetic) noise and inhomogeneous
broadening. Then, the first strong (noisy) driving field suppresses the fast noise of the environment.
Additionally, the robust DD sequences in the dressed basis compensate for the relatively slower noise of the
driving field. MDD also allows the first driving field (and its noise) to be stronger in comparison to double-drive concatenated CDD as the robust phased DD sequences typically have a wider error compensation range due to the phases of the pulses. MDD can also allow for faster quantum gates implementation than concatenated double-drive CDD as the gates can be implemented in the dressed basis of the first driving field and be embedded in the pulsed DD sequence.

We note that the limit in MDD performance is determined by the effect of the combined noise filter due to
the first continuous field and the robust phased sequence in the dressed basis. Similarly to standard concatenated CDD, MDD cannot compensate frequency components of the noise $\delta(t)$, which are faster than the amplitude of the first driving field, which is usually feasible. Furthermore, the application of the first continuous field with MDD leads to a reduction of the sensed field amplitude in the dressed basis by a factor of two in comparison to simple pulsed DD. Nevertheless, more advanced sensing protocols might be able to overcome this limitation \cite{StarkSciRep2018}.%, similarly to the improved continuous double-drive protocol in \cite{CohenFP2017}.

Finally, the optimum MDD protocol would depend on the characteristics of the dressed basis noise, i.e., the remnant environmental (magnetic) noise after the first driving field, which acts as a filter, and the amplitude noise of both fields. %The optimum robust phased sequence will depend on the specific noise spectrum with longer sequences allowing better compensation of pulse errors and slow amplitude noise. However, the total sequence duration should be short in comparison to the correlation time of the dressed basis noise (filtered by the first continuous field) to maintain appropriate phase relations between the pulses.
The availability of both pulsed and continuous MDD protocols allows for additional flexibility and fine-tuning to the environment. Pulsed MDD would be preferable when the zero frequency component of the amplitude noise in the first driving field is very high but its frequency range is limited, e.g., due to large inhomogeneous broadening. This allows for application short $\pi$ pulses with a large bandwidth and long pulse separation to minimize power input. Continuous MDD is expected to perform better with wider noise spectra (due to the high repetition rate of the $\pi$ pulses) and limited zero frequency noise (due to the lower Rabi frequency to maintain the same power input). Nevertheless, both pulsed and continuous MDD protocols allow for robust performance for a wider range of parameters, as compared to available schemes in the pulsed and continuous regime.

%%%%%%%%%%%%%%%%%%%%%%%%%%%%%%%%%%%%%%%%%%%%%%%%%%%%%%%%%%%%%%%%%%%%%%%%%%%%%%%%%%%%%%%%%%%%%%%%%%%%%%%%%%%%%%%%%%%%%%%%%%%%%%%
%%%%%%%%%%%%%%%%%%%%%%%%%%%%%%%%%%%%%%%%%%%%%%%%%%%%%%%%%%%%%%%%%%%%%%%%%%%%%%%%%%%%%%%%%%%%%%%%%%%%%%%%%%%%%%%%%%%%%%%%%%%%%%%
\section{Conclusion}\label{Section:Conclusion}
%%%%%%%%%%%%%%%%%%%%%%%%%%%%%%%%%%%%%%%%%%%%%%%%%%%%%%%%%%%%%%%%%%%%%%%%%%%%%%%%%%%%%%%%%%%%%%%%%%%%%%%%%%%%%%%%%%%%%%%%%%%%%%%
%%%%%%%%%%%%%%%%%%%%%%%%%%%%%%%%%%%%%%%%%%%%%%%%%%%%%%%%%%%%%%%%%%%%%%%%%%%%%%%%%%%%%%%%%%%%%%%%%%%%%%%%%%%%%%%%%%%%%%%%%%%%%%%

We introduced theoretically the idea for mixed dynamical decoupling (MDD), where we combine continuous dynamical decoupling (CDD) with robust phased DD sequences. % The main idea of MDD is to combine continuous DD with a strong (noisy) field and a robust sequence of (time-separated) phased pulses to protect from both field and environmental noise.
Specifically, we applied the same first driving field as with standard CDD, but we replaced the second continuous driving field with a robust sequence of phased pulses. We showed that MDD
with the same (or slightly lower) average power of the driving fields
outperforms standard concatenated CDD in realistic numerical simulations for DD in NV centers in diamond. Moreover, MDD with two noisy fields outperforms the concatenated CDD sequence even when we assume that the second driving field is noiseless for concatenated CDD. The reason is that MDD compensates better both the noise in the second driving field, as well as (part of the) error due to the noisy first driving field, which is not suppressed even by an ideal second field. Finally, we also demonstrated how our technique can be utilized for improved sensing. % in the pulsed and continuous regime.

As MDD is effectively a combined noise filter due to continuous and pulsed DD with robust phased pulses, its improved performance allows for higher efficiency and wider range of applications as it combines the advantages of both schemes. Examples for applications include improved quantum memories and sensing, e.g., in systems where the correlation time of the environment is too short to apply robust sequences of pulses directly in the bare basis, or faster gate implementations in comparison to double-drive CDD.

\acknowledgments
A. R. acknowledges the support of ERC grant QRES, project No. 770929, grant agreement No. 667192 (Hyperdiamond), the MicroQC, the ASTERIQS and the DiaPol projects. F. J. acknowledges the support of ERC, BMBF, DFG, Landesstiftung BW, and VW Stiftung.

\appendix
%%%%%%%%%%%%%%%%%%%%%%%%%%%%%%%%%%%%%%%%%%%%%%%%%%%%%%%%%%%%%%%%%%%%%%%%%%%%%%%%%%%%%%%%%%%%%%%%%%%%%%%%%%%%%%%%%%%%%%%%%%%%%%%
%%%%%%%%%%%%%%%%%%%%%%%%%%%%%%%%%%%%%%%%%%%%%%%%%%%%%%%%%%%%%%%%%%%%%%%%%%%%%%%%%%%%%%%%%%%%%%%%%%%%%%%%%%%%%%%%%%%%%%%%%%%%%%%
\section{Analytical calculation of fidelity errors}\label{Section:Fidelity_errors}
%%%%%%%%%%%%%%%%%%%%%%%%%%%%%%%%%%%%%%%%%%%%%%%%%%%%%%%%%%%%%%%%%%%%%%%%%%%%%%%%%%%%%%%%%%%%%%%%%%%%%%%%%%%%%%%%%%%%%%%%%%%%%%%
%%%%%%%%%%%%%%%%%%%%%%%%%%%%%%%%%%%%%%%%%%%%%%%%%%%%%%%%%%%%%%%%%%%%%%%%%%%%%%%%%%%%%%%%%%%%%%%%%%%%%%%%%%%%%%%%%%%%%%%%%%%%%%%

\begin{figure}[t!]
\includegraphics[width=\columnwidth]{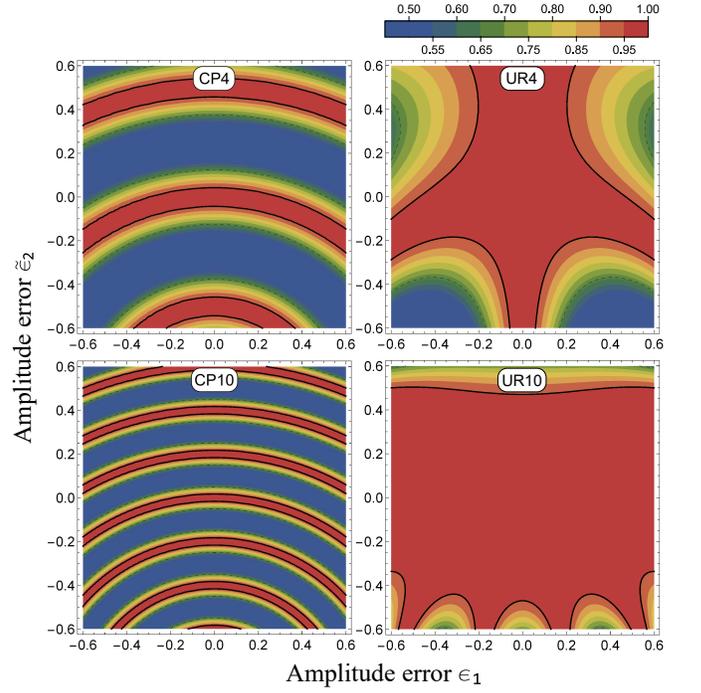}
\caption{(color online)
Fidelity vs. variation in the amplitude errors $\widetilde{\epsilon_1}$ and $\epsilon_2$ of the first and second driving fields for: (top, left) the Carr-Purcell (CP) sequence with a pulse area $A=4\pi$ and (top, right) the respective UR4 robust sequence with the same target pulse area; (bottom, left) the CP sequence with a pulse area $A=10\pi$ and (bottom, right) the respective UR10 robust sequence with the same target pulse area. The dashed (bold, solid) contour corresponds to $F=0.67$ ($F=0.95$). The fidelity calculation is based on the Hamiltonian in Eq. \eqref{Eq:HI2phi} in the approximation of static amplitude noise, i.e., $\epsilon_{k}(t)=\epsilon_{k},~k=1,2$. We note that $\widetilde{\epsilon}_1\equiv \epsilon_{1}\Omega_1/\Omega_2$ is the rescaled amplitude error of the first driving field. All sequences assumed a zero pulse separation $\tau=0$.
}
\label{Fig6_plots}
\end{figure}

In order to compare robustness, we give examples for the analytical calculation of the error in the fidelity of several DD sequences in this section. The calculations are based on the Hamiltonian in Eq. \eqref{Eq:HI2phi}. We make the approximation that the amplitude noise is static, i.e., $\epsilon_{k}(t)=\epsilon_{k},~k=1,2$ and assume a zero pulse separation $\tau=0$ for simplicity and to obtain intuition. We also define $\widetilde{\epsilon}_1\equiv \epsilon_{1}\Omega_1/\Omega_2$. Then, the Hamiltonian in Eq. \eqref{Eq:HI2phi} is transformed to
\begin{equation}\label{Eq:HI2phi_rescaled}
\widetilde{H}_{I2}(\phi)=\frac{\Omega_2}{2}\left[\widetilde{\epsilon}_1\sigma_{x}+(1+\epsilon_2)\left(\cos{\phi}\sigma_{y}+\sin{\phi}\sigma_{z}\right)\right].
\end{equation}
We then use Eq. \eqref{Eq:Fid_definition} and calculate the fidelity with double-drive CDD ($\phi=0$), which takes the form
\begin{equation}\label{Eq:Fid_CCDD}
F=\frac{1}{3}\left(2+\cos{A\sqrt{\widetilde{\epsilon}_1^2+(1+\epsilon_2)^2}} \right),
\end{equation}
where $A\equiv \Omega_2 t$ is the target pulse area of the second drive, which we assume to be continuous. It is evident that the error in the fidelity accumulates with time for non-zero values of $\widetilde{\epsilon}_1$ and $\epsilon_2$ as the error in the effective pulse area $\Omega_2 t\sqrt{\widetilde{\epsilon}_1^2+(1+\epsilon_2)^2}$ increases. In the case of a noiseless second drive ($\epsilon_2=0$), the fidelity becomes
\begin{align}\label{Eq:Fid_CCDD_noiseless_2}
F&=\frac{1}{3}\left(2+\cos{A\sqrt{\widetilde{\epsilon}_1^2+1}}\right)\approx \frac{1}{3}\left(2+\cos{A(1+\widetilde{\epsilon}_1^2/2)}  \right)\notag\\
\end{align}
where we assumed that $\widetilde{\epsilon}_1\ll 1$ for this approximation. Thus, the error in the pulse area from the target one due to the noisy first drive is $A \widetilde{\epsilon}_1^2/2=\epsilon_1^2\Omega_1^2 t/2\Omega_2\sim \epsilon_1^2\Omega_1^2/\Omega_2$. As double-drive CDD requires $\Omega_1\gg \Omega_2$ for RWA to be valid, the effect of the term $\epsilon_1^2\Omega_1^2/\Omega_2$ due the noisy first drive often cannot be neglected. Usually, we perform measurements when $A=2\pi m,m\in N$, i.e., $F=1$ in the absence of noise. Then, the error in the fidelity is given by
\begin{align}\label{Eq:Fid_CCDD_noiseless_2_continued}
\varepsilon\equiv 1-F&\approx \frac{A^2 \widetilde{\epsilon}_1^4}{24}+O(\widetilde{\epsilon}_1 ^{6})=\frac{m^2\pi^2 \widetilde{\epsilon}_1^4}{6}+O(\widetilde{\epsilon}_1 ^{6}),
\end{align}
where the last approximation is valid when $A \widetilde{\epsilon}_1^2\lesssim 0.1\pi$.
In the case when the first field is noiseless, the error in the fidelity with respect to noise in the second field
%\begin{align}\label{Eq:Fid_CCDD_noiseless_2_continued2}
%F &= \frac{1}{3}\left[2+\cos{(A(1+\epsilon_2))}\right],
%\end{align}
% Usually, we perform measurements
at a target pulse area $A=2\pi m,m\in N$ is given by
% , when the error in the fidelity takes the form
\begin{align}
\varepsilon &=\frac{2}{3}\sin^2(A\epsilon_2/2)\approx \frac{A^2 \widetilde{\epsilon}_2^2}{6},
\end{align}
where the last approximation is valid when $A \widetilde{\epsilon}_1^2\lesssim 0.1\pi$.
When both fields are noisy the error in the fidelity is
\begin{align}
\varepsilon &\approx \frac{A^2 \widetilde{\epsilon}_1^4}{24}+\left(\frac{1}{3}+\frac{\epsilon_1^2}{4}\right)\frac{A^2 \widetilde{\epsilon}_1^2 \epsilon_2}{2}+\left(1-\widetilde{\epsilon}_1^2%+\frac{A^2(A^2-12)}{8}\widetilde{\epsilon}_1^4
\right)\frac{A^2 \epsilon_2^2}{6},
\end{align}
%+\left(\frac{A^2\epsilon_1^4}{8}+\frac{A^2\epsilon_1^2}{6}+O(\widetilde{\epsilon}_1^6}+O(\widetilde{\epsilon}_1^6)\right)
where the lowest order mixed error term is $\sim A^2 \widetilde{\epsilon}_1^2 \epsilon_2$. The variation in fidelity in the presence of errors in both amplitudes is shown in Fig. \ref{Fig6_plots} (left column). We note that concatenated double drive case corresponds to the standard Carr-Purcell (CP) sequence with pulse separation $\tau=0$, which we use as a label in the figure.

Next, we show how we can improve performance by using phased sequences of pulses in the dressed basis. We assume for simplicity and without loss of generality that the second drive is continuous but we apply phase shifts at particular times, such that the target pulse area of each time period with a constant phase (which we term a pulse) is $A=\Omega_2 T=\pi$. The dynamics of the system during the first pulse from the second drive is described by the propagator $U_{I2}(\phi_1)=\exp{\left(-\i\widetilde{H}_{I2}(\phi_1)T\right)}$. Then, the propagator of a DD sequence of $n$ pulses, where the $k$-th pulse is phase shifted by $\phi_{k}$, takes the form
\begin{equation}\label{Eq:UnI2}
U_{I2}^{(n)}=U_{I2}(\phi_n)\dots U_{I2}(\phi_2)U_{I2}(\phi_1),
\end{equation}
where $\phi_1,\dots,\phi_{\text{n}}$ are free control parameters. The DD sequence can be repeated $N$ times for decoupling during the whole storage time.

%%%%%%%%%%%%%%%%%%%%%%%%%% FIGURE 5 %%%%%%%%%%%%%%%%%%%%%%%%%%%%%
%%%%%%%%%%%%%%%%%%%%%%%%%% FIGURE 5 %%%%%%%%%%%%%%%%%%%%%%%%%%%%%

First, we give an example for the UR4 sequence, shown in Fig. \ref{Fig2:CP_UR4}. We again use Eq. \eqref{Eq:Fid_definition} to calculate the error in the fidelity, which in case of a noiseless second drive ($\epsilon_2=0$) takes the form
\begin{align}\label{Eq:Fid_UR4}
\varepsilon_{\text{UR4}}&=\frac{4}{3y^4}\left(y^2-1\right)\sin ^2(\pi  y)  \notag\\
&\times \left(y^2+\left(y^2-1\right)\cos (2 \pi  y)+1\right),
\end{align}
where $y\equiv \sqrt{1+\widetilde{\epsilon}_1^2}$. This expression can be approximated to
\begin{equation}\label{Eq:Fid_UR4_approx}
\varepsilon_{\text{UR4}}\approx \frac{2\pi^2}{3} \widetilde{\epsilon}_1 ^{6}+O(\widetilde{\epsilon}_1 ^{8}),
\end{equation}
where we assumed that $\widetilde{\epsilon}_1\ll 1$ for the approximation. It is evident that the error in the fidelity for the UR4 sequence is improved significantly as it is proportional to $\sim \widetilde{\epsilon}_1^6$, while the error with standard double drive is $\sim \widetilde{\epsilon}_1^4$ (note that $\widetilde{\epsilon}_1\ll 1$).
We note that the UR4 sequence is insensitive to variation in the amplitude of the second drive when the first field is noiseless, i.e., then, the errors due to $\epsilon_2$ are fully compensated. When both fields are noisy, the error in the fidelity can be approximated by
\begin{equation}\label{Eq:Fid_UR4_approx_both}
\varepsilon_{\text{UR4}}\approx \frac{2\pi^2}{3} \widetilde{\epsilon}_1 ^{6}+\frac{8\pi^2}{3} \widetilde{\epsilon}_1 ^{4}\epsilon_2+\frac{8\pi^2(1-4\widetilde{\epsilon}_1 ^{2})}{3}\widetilde{\epsilon}_1 ^{2}\epsilon_2^2,
\end{equation}
where the lowest order mixed error terms are $\sim \widetilde{\epsilon}_1 ^{4}\epsilon_2$ and $\sim \widetilde{\epsilon}_1 ^{2}\epsilon_2^2$ and are much smaller than the lowest order mixed term for continuous double-drive with a constant zero phase $\sim \widetilde{\epsilon}_1^2 \epsilon_2$. The fidelity variation in the presence of errors in both amplitudes for the UR4 sequence is shown in Fig. \ref{Fig6_plots} (top, right). As UR4 contains four pulses, its total pulse area is ideally $4\pi$, which allows for direct comparison with the CP sequence with the same target pulse area, termed CP4, in Fig. \ref{Fig6_plots} (top, left). It is evident that the UR4 sequence allows for a much broader range of amplitude errors, within which the fidelity remains high.

Higher order robust DD sequences can improve performance even further. For example, the UR10 sequence \cite{GenovPRL2017}, which we use in the numerical simulations has an error in the fidelity $\varepsilon_{\text{UR10}}\sim O(\widetilde{\epsilon}_1^{12})$ for noiseless second drive
and $\varepsilon_{\text{UR10}}\sim O(\epsilon_2^{10})$ for a noiseless first drive. It also improves performance in the presence of noise in both driving fields, which is demonstrated in in Fig. \ref{Fig6_plots} (bottom, right). As the UR10 sequence consists of ten pulses with a target total pulse area of $10\pi$, we can compare directly with the CP sequence with the same pulse area, termed CP10, in Fig. \ref{Fig6_plots} (bottom, left). It is evident that the UR10 phased sequence allows for a much broader range of amplitude errors, within which the fidelity remains high.
Errors can be reduced to even higher order by applying longer robust sequences of phased pulses as long as the errors remain approximately the same during a sequence. This is necessary as the error of one pulse is compensated by the subsequent pulses in a sequence. Thus, the correlation time of the amplitude noise should be long enough, so that the errors in the amplitudes of both fields do not change significantly during a single DD sequence, which is usually feasible.
%We note that other robust phased sequences can also be applied, e.g., the widely used XY4, XY8 or KDD sequences \cite{Suter16RMP,DegenRMP2017} and the optimum sequence will depend on the specific environment.

%%%%%%%%%%%%%%%%%%%%%%%%%%%%%%%%%%%%%%%%%%%%%%%%%%%%%%%%%%%%%%%%%%%%%%%%%%%%%%%%%%%%%%%%%%%%%%%%%%%%%%%%%%%%%%%%%%%%%%%%%%%%%%%
%%%%%%%%%%%%%%%%%%%%%%%%%%%%%%%%%%%%%%%%%%%%%%%%%%%%%%%%%%%%%%%%%%%%%%%%%%%%%%%%%%%%%%%%%%%%%%%%%%%%%%%%%%%%%%%%%%%%%%%%%%%%%%%
\section{Numerical Simulation}\label{Section:Numerics}
%%%%%%%%%%%%%%%%%%%%%%%%%%%%%%%%%%%%%%%%%%%%%%%%%%%%%%%%%%%%%%%%%%%%%%%%%%%%%%%%%%%%%%%%%%%%%%%%%%%%%%%%%%%%%%%%%%%%%%%%%%%%%%%
%%%%%%%%%%%%%%%%%%%%%%%%%%%%%%%%%%%%%%%%%%%%%%%%%%%%%%%%%%%%%%%%%%%%%%%%%%%%%%%%%%%%%%%%%%%%%%%%%%%%%%%%%%%%%%%%%%%%%%%%%%%%%%%

We perform a numerical simulation where we apply DD in a two-state system, which is subject to magnetic noise $\delta(t)$ and uncorrelated power fluctuations of the driving fields. The parameters of the noise have the characteristics for typical experiments in NV centers, as described in \cite{CaiNJP2012,AharonNJP2016}. Specifically, the noise $\delta(t)$ is modelled as an Ornstein-Uhlenbeck (OU) process \cite{UhlenbeckRMP1945,GillespieAJP1996} with a zero expectation value $\langle \delta(t)\rangle = 0$, correlation function $\langle \delta(t)\delta(t^{\prime})\rangle =(1/2)c\widetilde{\tau}\exp{(-\gamma|t-t^{\prime}|)}$, where $c$ is a diffusion constant and $\widetilde{\tau}=1/\gamma$ is the correlation time of the noise. The OU process is implemented with an exact algorithm \cite{GillespieAJP1996}
\begin{equation}\label{Eq:OU_noise}
\delta(t+\Delta t)=\delta(t)\e^{-\frac{\Delta t}{\widetilde{\tau}}}+\widetilde{n}\sqrt{\frac{c\widetilde{\tau}}{2}\left(1-\e^{-\frac{2\Delta t}{\widetilde{\tau}}}\right)},
\end{equation}
where $\widetilde{n}$ is a unit Gaussian random number. The correlation time of the noise is $\widetilde{\tau}=25 \mu$s with a diffusion constant $c\approx 4/(T_{2}^{\ast} \widetilde{\tau})$, where $T_{2}^{\ast}=3 \mu$s \cite{CohenFP2017}.  The driving fluctuations are also modelled by uncorrelated OU processes with the same correlation time $\tau_{\Omega}=500 \mu$s and a relative amplitude error $\delta_{\Omega}=0.005$ with the corresponding diffusion constant $c_{\Omega}=2\delta_{\Omega_{i}}^2\Omega_{i}^2/\tau_{\Omega},~i=1,2$.

We note that we model the system in the first interaction basis with RWA ($\Omega_1\ll\omega_0$) with the Hamiltonian given by
\begin{equation}\label{H:fid_sims}
\widetilde{H}_{I1}(\phi,t)=H_{I1}+\Omega_2(t) (1+\epsilon_2(t))\sigma_{y}\cos{(\Omega_1 t+\phi(t))},
\end{equation}
where $\Omega_2(t)$ and $\phi(t)$ are assumed time-dependent piecewise functions, e.g., $\Omega_2(t)=\Omega_2$ when we apply a DD pulse with the second field in the first order interaction basis and $\Omega_2(t)=0$ in case of no pulse from the second field. We choose not to work in the second order interaction basis, as defined in Eq. \eqref{Eq:HI2phi}, as $\delta(t)$ can often not be neglected. This is especially true when the power of the first driving field is not much greater than zero frequency power spectrum component of the $\delta(t)$ noise, e.g., in case of large inhomogenous broadening. Additionally, we do not apply the RWA in the second order interaction basis ($\Omega_2\ll\Omega_1$). This approach expands significantly the parameter range when our numerical simulation is applicable.

%%%%%%%%%%%%%%%%%%%%%%%%%% FIGURE 3 %%%%%%%%%%%%%%%%%%%%%%%%%%%%%
\begin{figure}[t!]
\includegraphics[width=\columnwidth]{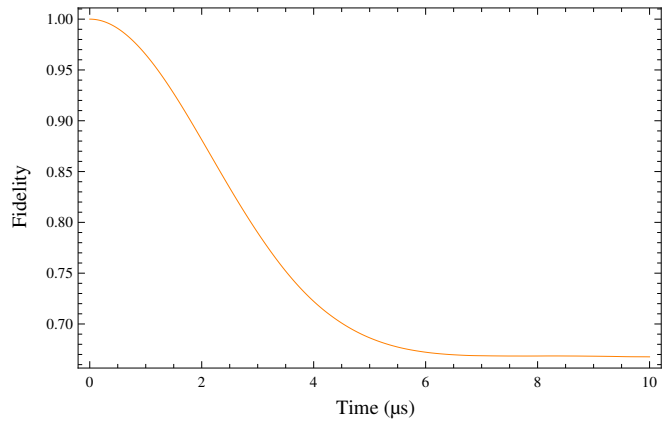}
\caption{(color online)
Pure dephasing simulation shows the expected decay of the fidelity to the quantum limit of $0.667$. %The decay time $T_{2}^{\ast}\approx 3 \mu$s corresponds to previous simulations for NV centers \cite{CohenFP2017}.
}
\label{Fig3:Pure_dephasing}
\end{figure}
%%%%%%%%%%%%%%%%%%%%%%%%%% FIGURE 3 %%%%%%%%%%%%%%%%%%%%%%%%%%%%%

Then, we calculate numerically the propagator
\begin{equation}
\widetilde{U}_{I1}(t,t_0)=\T \exp{\left(-\i\int_{t_0}^{t}\widetilde{H}_{I1}(t^{\prime})\d t^{\prime}\right)}
\end{equation}
for the particular noise realisation of $\delta(t)$, $\epsilon_1(t)$ and $\epsilon_2(t)$ and the chosen DD sequence. We use a time-discretization with a time step of $10$ ns, which is comparable to the resolution of many standard arbitrary wave-form generators. We note that the OU noise characteristics are not affected by this choice of $\Delta t$, as Eq. \eqref{Eq:OU_noise} is exact. We can then make use of the calculated $\widetilde{U}_{I1}(t,t_0)$ and obtain the time evolution of the fidelity
\begin{equation}
F(t)=\frac{1}{3}\sum_{k=x,y,z} \text{Tr} \left( \widetilde{U}_{I1}(t,t_0)\rho_{k}\left(\widetilde{U}_{I1}(t,t_0)\right)^\dagger\rho_{k} \right)
\end{equation}
for the particular noise realization. The expected average fidelity is calculated by performing the simulation $2500$ times for different noise realizations.
Our simulation agrees well with previous numerical results for NV centers \cite{AharonNJP2016}. For example, Fig. \ref{Fig3:Pure_dephasing}) shows that our simulation exhibits the expected dephasing time of $T_{2}^{\ast}\approx 3 \mu$s.

The MDD sensing simulations in Figs. \ref{Fig4pulsed} and \ref{Fig5_UR10cw} use the same noise characteristics and time-discretization as above. The Hamiltonian is the same as in Eq. \eqref{H:fid_sims} plus a term for the sensed field:
\begin{align}
\widetilde{H}_{1,s}(t)&=\frac{\delta(t)}{2}\sigma_{z}+\frac{\Omega_1}{2} (1+\epsilon_1(t))\sigma_{x}\\
&+\Omega_2(t) (1+\epsilon_2(t))\sigma_{y}\cos{(\Omega_1 t+\phi(t))}\notag\\
&+\frac{g}{2}\left(\sigma_{x}\cos{(\Delta t+\xi)}+\sigma_{y}\sin{(\Delta t+\xi)}\right).\notag
\end{align}
Again, we calculate numerically the propagator
\begin{equation}
\widetilde{U}_{1,s}(t,t_0)=\T \exp{\left(-\i\int_{t_0}^{t}\widetilde{H}_{1,s}(t^{\prime})\d t^{\prime}\right)}
\end{equation}
for the particular noise realisation of $\delta(t)$, $\epsilon_1(t)$ and $\epsilon_2(t)$ and the chosen DD sequence. We then make use of the calculated $\widetilde{U}_{1,s}(t,t_0)$ and obtain the time evolution of the density matrix
\begin{equation}
\rho(t)=\widetilde{U}_{1,s}(t,t_0)\rho(t_0)\widetilde{U}^\dagger_{1,s}(t,t_0),
\end{equation}
where $\rho(t_0)=\rho_{z}\equiv(\sigma_{0}+\sigma_{z})/2$ is the initial density matrix, which corresponds to preparation of the system in the ground state. The expected density matrix $\overline{\rho}(t)$ is calculated by performing the simulation $2500$ times for different noise realizations and averaging the result.

Figures \ref{Fig4pulsed} and \ref{Fig5_UR10cw} show the expected population of the ground state in the bare basis at times $m(T+\tau)$ when the noise is ideally refocused. We note that simulation results for sensing can be corrected for the expected population inversion and phase evolution. For example, this is necessary when we measure at times when we have applied an odd number of $\pi$ pulses by the second driving field and the population is inverted in the bare basis. In order to do this we calculate the propagator for evolution without noise and without a sensed field
\begin{align}
&\widetilde{U}^{(0)}_{1,s}(t,t_0)=\T \exp{\left(-\i\int_{t_0}^{t}\widetilde{H}^{(0)}_{1,s}(t^{\prime})\d t^{\prime}\right)},\text{ where}\notag\\
&\widetilde{H}^{(0)}_{1,s}(t)=\frac{\Omega_1}{2}\sigma_{x}+\Omega_2(t)\sigma_{y}\cos{(\Omega_1 t+\phi(t))}
\end{align}
and obtain the corrected density matrix
\begin{equation}
\widetilde{\overline{\rho}}(t)=\left(\widetilde{U}^{(0)}_{1,s}(t,t_0)\right)^\dagger\overline{\rho}(t)\widetilde{U}^{(0)}_{1,s}(t,t_0),
\end{equation}
The simulations show the corrected population in the ground state $\widetilde{\overline{\rho}}_{11}(t)$. We note that no correction is needed at times, which correspond to complete DD sequences and when $\Omega_{1} t=0~(\text{mod}~2\pi)$. For example, no corrections were applied to the simulation results in Fig. \ref{Fig5_UR10cw}, which were taken at an interval of $80~\mu$s, corresponding to the duration of the UR10 sequence.

%%%%%%%%%%%%%%%%%%%%%%%%%%%%%%%%%%%%%%%%%%%%%%%%%%%%%%%%%%%%%%%%%%%%%%%%%%%%%%%%%%%%%%%%%%%%%%%%%%%%%%%%%%%%%%%%%%%%%%%%%%%%%%%
%%%%%%%%%%%%%%%%%%%%%%%%%%%%%%%%%%%%%%%%%%%%%%%%%%%%%%%%%%%%%%%%%%%%%%%%%%%%%%%%%%%%%%%%%%%%%%%%%%%%%%%%%%%%%%%%%%%%%%%%%%%%%%%
\section{Signal phase selectivity of the continuous MDD protocol}\label{Section:MDD_phase_selectivity}
%%%%%%%%%%%%%%%%%%%%%%%%%%%%%%%%%%%%%%%%%%%%%%%%%%%%%%%%%%%%%%%%%%%%%%%%%%%%%%%%%%%%%%%%%%%%%%%%%%%%%%%%%%%%%%%%%%%%%%%%%%%%%%%
%%%%%%%%%%%%%%%%%%%%%%%%%%%%%%%%%%%%%%%%%%%%%%%%%%%%%%%%%%%%%%%%%%%%%%%%%%%%%%%%%%%%%%%%%%%%%%%%%%%%%%%%%%%%%%%%%%%%%%%%%%%%%%%

In order to analyze the signal phase selectivity of the continuous MDD scheme it proves useful to consider the propagator, which determines the evolution of the system from a starting time $t_0$ to a later time $t$ in the basis of the Hamiltonian in Eq. \eqref{Eq:H_frame_4s}:
\begin{align}\label{Eq:U_frame_4s}
\widetilde{U}_{\text{4,s}}&(t,t_0)=\exp[-\i \widetilde{H}_{\text{4,s}}(t-t_0)]\\
&=\sigma_0\cos{\left(\widetilde{\Theta}/2\right)}+\i\sin{\left(\widetilde{\Theta}/2\right)}\left[\sigma_{z}\sin{(\xi)}-\sigma_{x}\cos{(\xi)}\right],\notag
\end{align}
where $\widetilde{\Theta}=g(t-t_0)/2$. Then, it is straightforward to determine also the propagator in the basis of the Hamiltonian in Eq. \eqref{Eq:H_frame_3s}:
\begin{align}\label{Eq:U_frame_3s}
\widetilde{U}_{\text{3,s}}&(t,t_0)=R_4^{\dagger}(t-t_0)\widetilde{U}_{\text{4,s}}(t,t_0)R_4(t-t_0)\\
&=\sigma_0\cos{\left(\widetilde{\Theta}/2\right)}+\i\sin{\left(\widetilde{\Theta}/2\right)}\left[\sigma_{z}\sin{(\widetilde{\xi})}-\sigma_{x}\cos{(\widetilde{\xi})}\right],\notag
\end{align}
where $\widetilde{\xi}\equiv\xi+\Omega_2 (t-t_0)$, i.e., the propagator $\widetilde{U}_{\text{3,s}}(t,t_0)$ is obtained from $\widetilde{U}_{\text{4,s}}(t,t_0)$ by taking $\xi\to\widetilde{\xi}$. Finally, the propagator in the phase-independent basis of the Hamiltonian in Eq. \eqref{H_sensing} is given by
\begin{align}\label{Eq:U_frame_2s}
\widetilde{U}_{\text{2,s}}&(\phi,t,t_0)=R_3^{\dagger}(\phi)\widetilde{U}_{\text{3,s}}(t,t_0)R_3(\phi),
\end{align}
where $\phi$ is the phase of the second field during the time period from $t_0$ to $t$. Then, the propagator of a sequence of two pulses for time evolution from $t_0$ to $t$, where we apply a phase change from $\phi_1$ to $\phi_2$ at time $t_1\in (t_0,t)$ is given by
\begin{align}\label{Eq:U_2p}
\widetilde{U}_{\text{2,s}}^{(\phi_1,\phi_2)}&=\widetilde{U}_{\text{2,s}}(\phi_2,t,t_1)\widetilde{U}_{\text{2,s}}(\phi_1,t_1,t_0)\\
&=R_3^{\dagger}(\phi_2)\widetilde{U}_{\text{3,s}}(t,t_1)R_3(\Delta \phi)\widetilde{U}_{\text{3,s}}(t_1,t_0)R_3(\phi_1),\notag
\end{align}
where $\Delta\phi=\phi_2-\phi_1$. The effect of the phase change depends on the commutator
\begin{align}\label{Eq:Commutator}
\left[R_3(\Delta\phi),\widetilde{U}_{\text{3,s}}(t_1,t_0)\right]=2\i\sigma_{y}\sin{(\widetilde{\Theta}_1/2)}\sin{(\Delta\phi/2)}\sin{(\widetilde{\xi}_1)},
\end{align}
where $\widetilde{\Theta}_1=g(t_1-t_0)/2$ is the rotation angle due to the sensed field during the time evolution between $t_0$ and $t_1$, while $\widetilde{\xi}_1=\xi+\Omega_2(t_1-t_0)$ depends on the initial phase of the sensed field $\xi$ at time $t_0$ and the pulse area of the second driving field from $t_0$ to $t_1$. When the commutator is zero, the propagator becomes
\begin{align}\label{Eq:U_2p_0}
\widetilde{U}_{\text{2,s}}^{(\phi_1,\phi_2)}&=R_3^{\dagger}(\phi_2)\widetilde{U}_{\text{3,s}}(t,t_0)R_3(\phi_2),\notag
\end{align}
which is the same as the propagator when we apply a second drive with a constant phase $\phi_2$ during the whole interaction from $t_0$ to $t$.

It is evident that when a sensed field is present ($g\ne 0$), the commutator is zero for any $\xi$ in the special case when $\Delta\phi=0~(\text{mod}~2\pi)$, which corresponds to standard double-drive CDD. In the more general case of arbitrary phase changes $\Delta\phi$, the commutator in Eq. \eqref{Eq:Commutator} will be zero when
\begin{equation}\label{Eq:Select_cond_MDD}
\widetilde{\xi}_1=\xi+\Omega_2(t_1-t_0)=0~(\text{mod}~\pi).
\end{equation}
For example, this condition is satisfied for an initial phase $\xi=0$ and $t_1-t_0=T=\pi/\Omega_2$, i.e., when the center of first $\pi$ pulse with a phase $\phi_1$ corresponds to the time when the signal $g(t)$ changes its sign. This is exactly the same condition as with pulsed MDD. On the contrary, the commutator will be non-zero for the signal component with an initial phase $\xi=\pi/2$. Then, this component of the signal will experience more complex time-evolution and will usually be suppressed by the robust MDD sequence. Thus, the possibility to apply arbitrary phase changes in the MDD protocol leads to selectivity with respect to the initial phase of the signal and the selectivity condition is the same as for the pulsed MDD scheme.

Next, we note that the same analysis applies for any subsequent phase changes in the MDD protocol. As these occur at time intervals of $T=\pi/\Omega_2$, the phase $\widetilde{\xi}_{k}$ at the time of the $k$-th phase change will be $\widetilde{\xi}_{k}=\widetilde{\xi}_{1}~(\text{mod}~\pi)$. Thus, if the first commutator in Eq. \eqref{Eq:Commutator} is zero, all subsequent commutators will also be zero and the signal will be filtered through the MDD sequence. %similarly to the standard double-drive CDD.
Finally, we note that we can choose the ``good'' initial phase $\xi$ of the signal by shifting the whole MDD sequence in time. We only need to change the duration of the first pulse period to satisfy $\widetilde{\xi}_1=\xi+\Omega_2(t_1-t_0)=0$ and apply the subsequent phase changes at intervals of $T$.

In summary, the continuous MDD protocol allows for arbitrary phase changes of the second driving field at times, separated by $T=\pi/\Omega_2$ and has the same signal phase selectivity condition as the pulsed MDD protocol. In contrast, concatenated double-drive CDD works efficiently for any initial phase $\xi$ of the sensed signal field. Nevertheless, continuous MDD allows for the application of robust phased sequences of pulses in the dressed basis with their improved noise-suppression characteristics.

\black

%%%%%%%%%%%%%%%%%%%%%%%%%%%%%%%%%%%%%%%%%%%%%%%%%%%%%%%%%%%%%%%%%%%%%%%%%%%%%%%%%%%%%%%%%%%%%%%%%%%%%%%%%%%%%%%%%%%%%%%%%%%%%%%%%%%
%%%%%%%%%%%%%%%%%%%%%%%%%%%%%%%%%%%%%%%%%%%%%%%%%%%%%%%%%%%%%%%%%%%%%%%%%%%%%%%%%%%%%%%%%%%%%%%%%%%%%%%%%%%%%%%%%%%%%%%%%%%%%%%%%%%
%%%%%%%%%%%%%%%%%%%%%%%%%%%%%%%%%%%%%%%%%%%%%%%%%%%%%%%%%%%%%%%%%%%%%%%%%%%%%%%%%%%%%%%%%%%%%%%%%%%%%%%%%%%%%%%%%%%%%%%%%%%%%%%%%%%
%%%%%%%%%%%%%%%%%%%%%%%%%%%%%%%%%%%%%%%%%%%%%%%%%%%%%%%%%%%%%%%%%%%%%%%%%%%%%%%%%%%%%%%%%%%%%%%%%%%%%%%%%%%%%%%%%%%%%%%%%%%%%%%%%%%
%%%%%%%%%%%%%%%%%%%%%%%%%%%%%%%%%%%%%%%%%%%%%%%%%%%%%%%%%%%%%%%%%%%%%%%%%%%%%%%%%%%%%%%%%%%%%%%%%%%%%%%%%%%%%%%%%%%%%%%%%%%%%%%%%%%

\end{document}